\documentclass[3p,twocolumn]{elsarticle}

\usepackage{amsfonts,amsmath,amssymb}
\usepackage{graphicx,color}

\def\blfootnote{\xdef\@thefnmark{}\@footnotetext}
\begin{document}

\begin{frontmatter}

\title{Heavy Quarkonium Dissociation by Thermal Gluons at Next-to-leading Order in the Quark-Gluon Plasma}

\author[a]{Shile Chen}
\author[a]{Min He}

\address[a]{Department of Applied Physics, Nanjing University of Science and Technology, Nanjing 210094, China}

\date{\today}

\begin{abstract}
Using the chromo-electric dipole coupling Hamiltonian from QCD multipole expansion, we derive the dissociation cross sections of heavy quarkonia by thermal gluons at next-to-leading order (NLO, also known as inelastic parton scattering dissociation) in the Quark-Gluon Plasma (QGP) in the framework of second order quantum mechanical perturbation theory. While suffering divergence (infrared and soft-collinear divergences) in vacuum, the cross sections thus derived become finite in the QGP as rendered by the finite thermal gluon masses. In contrast to the leading order (LO, also known as gluo-dissociation) counterparts rapidly dropping off with increasing incident gluon energy, the NLO cross sections exhibits finite value toward high energies because of new phase space being opened up. We then carry out a full calculation of the dissociation rates for various charmonia and bottomonia within a non-relativistic in-medium potential model. The NLO process is shown to dominate the dissociation rate toward high temperatures when the binding energies of heavy quarkonia become smaller relative to the Debye screening mass.

\end{abstract}

\begin{keyword}
Heavy Quarkonium \sep Quark Gluon Plasma \sep Ultrarelativistic Heavy-Ion Collisions
\PACS 25.75.Dw \sep 12.38.Mh \sep 25.75.Nq
\end{keyword}

\end{frontmatter}

%%%%%%%%%%%%%%%%%%%%%%%%%%%%%%%%%%%%%%%%%%%%%%%%%%%%%%%%
\section{Introduction}
\label{sec_intro}
%%%%%%%%%%%%%%%%%%%%%%%%%%%%%%%%%%%%%%%%%%%%%%%%%%%%%%%
It was first suggested in the seminal work by Matsui and Satz~\cite{Matsui:1986dk} that heavy quarkonium (bound states of heavy quark-antiquark $Q\bar{Q}$) dissociation, as a result of the screening of the binding color forces at short distances in a hot thermal bath, can be used as a probe of the deconfined state of nuclear matter known as Quark-Gluon Plasma (QGP). This picture was supported by the experimental observation of the sequential suppression of the $\Upsilon$ states at the LHC~\cite{Chatrchyan:2012lxa}, with the less bound excited $2S$ and $3S$ states being more suppressed than the most tightly bound ground $1S$ states.

Besides static screening, however, other mechanisms involving in particular collisions with plasma particles can lead to dynamical dissociation of the bound states. These collisional processes generate inelastic widths that enter the transport equations describing the evolution (yields, momentum spectra) of the bound states in the QGP~\cite{Rapp:2008tf,Rapp:2017chc,Zhao:2010nk,Du:2017qkv,Zhou:2014kka,Strickland:2011mw,Song:2011nu}. In this respect, the one that has been intensively studied in the past decades is the gluo-dissociation of heavy quarkonium $g+\Psi\rightarrow Q+\bar{Q}$ ($\Psi$ denotes a $Q\bar{Q}$ bound state)~\cite{Peskin:1979va,Kharzeev:1994pz,Wong:2001kn,Arleo:2001mp,Oh:2001rm,Brezinski:2011ju,Brambilla:2008cx,Liu:2013kkg,Chen:2017jje}, an analog of photo-dissociation of neutral atoms owing to dipole transition. In this process, an energetic thermal gluon from the medium is absorbed by the bound state, thereby the binding energy of the bound state is overcome, and the latter is dissociated into an unbound color octet $(Q\bar{Q})_8$. The cross section of this leading order (LO) scattering of thermal gluons off heavy quarkonium, when convoluted with the gluon distribution function, leads to a dissociation rate that decreases with increasing temperature, see {\it e.g.},~\cite{Chen:2017jje}, due to the fact that as the binding energy of the bound state gets lower, the peak of the LO cross section shifts toward smaller gluon energy corresponding to very small phase space~\cite{Grandchamp:2001pf}.

It was first noted in~\cite{Grandchamp:2001pf} that this artifact of the LO approximation could be removed by taking into account the next-to-leading order (NLO) inelastic scattering $g/q+\Psi\rightarrow g/q+Q+\bar{Q}$, and a ``quasi-free" scenario was proposed, in which the incident thermal gluon/quark is not absorbed but instead is scattered off the heavy quark $Q$ (or $\bar{Q}$) inside the bound state by exchanging another space-like gluon. The NLO scattering of light quarks/gluons off heavy quarkonium was then computed in the perturbative QCD approach in~\cite{Song:2005yd}. However, after various divergences were carefully handled, the resulting cross section appeared negative in some kinematic region~\cite{Song:2005yd}, obscuring the physical significance of the result. A substantial step toward the understanding of the NLO contribution to the heavy quarkonium dissociation rate was the realization that the effective potential entering the Schr\"{o}dinger equation obeyed by the real-time propagator of a $Q\bar{Q}$ pair develops an imaginary part because of Landau damping of the exchanged space-like gluon between $Q$ and $\bar{Q}$~\cite{Laine:2006ns} in the environment of QGP. This imaginary part of heavy quark potential, now already being confirmed by first-principle lattice QCD calculations~\cite{Burnier:2014ssa}, was also evaluated in~\cite{Beraudo:2007ky} and interpreted in terms of the collisions of $Q$ or $\bar{Q}$ within the bound state with hot bath constituents, thereby making closer connection to the ``quasi-free" scenario of the NLO inelastic scattering~\cite{Grandchamp:2001pf}. In the framework of potential non-relativistic QCD (pNRQCD) at finite temperature built upon the hierarchies of non-relativistic and thermal scales typical of heavy quarkonium in the QGP, the authors of~\cite{Brambilla:2008cx,Brambilla:2013dpa} were able to derive the same imaginary potential and further identify the LO dissociation rate as arising from singlet-to-octet breakup, and the NLO contribution as Landau damping.

In the present work, we revisit the NLO heavy quarkonium dissociation by thermal gluons in the QGP from the dynamical scattering point of view. We calculate the NLO inelastic scattering cross section of heavy quarkonium with thermal gluons in the framework of second-order quantum mechanical perturbation, using a color-electric dipole coupling effective interaction Hamiltonian derived from QCD multipole expansion~\cite{Yan:1980uh,Sumino:2014qpa}. The method adopted here allows for systematic incorporation of bound state wave function, thereby going beyond the ``quasi-free" approximation. In the following Sec.~\ref{sec_deriving_cross-section}, we demonstrate the derivation of the NLO dissociation cross section, using the $1S$ charmonium state $J/\psi$ as an example. The cross section thus derived diverges in vacuum with massless gluons. In Sec.~\ref{sec_in-medium_model}, we employ a nonrelativistic in-medium potential model and carry out a full calculation of the NLO dissociation cross sections and pertinent dissociation rates for various charmonia and bottomonia in the QGP; here the NLO cross sections become physical and finite as rendered by the thermal gluon masses. These NLO dissociation rates indeed take over from the LO counterparts toward high temperatures; that is, the artifact inherent in the LO approximation is removed. Finally, we briefly summarize and give an outlook in Sec.~\ref{sec_sum}.

%%%%%%%%%%%%%%%%%%%%%%%%%%%%%%%%%%%%%%%%%%%%%%%%%%%%%%%%%%%%%%%%
\section{Deriving the NLO dissociation Cross Section}
\label{sec_deriving_cross-section}
%%%%%%%%%%%%%%%%%%%%%%%%%%%%%%%%%%%%%%%%%%%%%%%%%%%%%%%%%%%%%%%%

%%%%%%%%%%%%%%%%%%%%%%%%%%%%%%%%%%%%%%%%%%%%%%%%%%%%%%%%%%%%%%%%
\subsection{The effective interaction Hamiltonian}
\label{ssec_effectiveHamiltonian}
%%%%%%%%%%%%%%%%%%%%%%%%%%%%%%%%%%%%%%%%%%%%%%%%%%%%%%%%%%%%%%%%

\begin{figure} [!t]
\includegraphics[width=1.45\columnwidth]{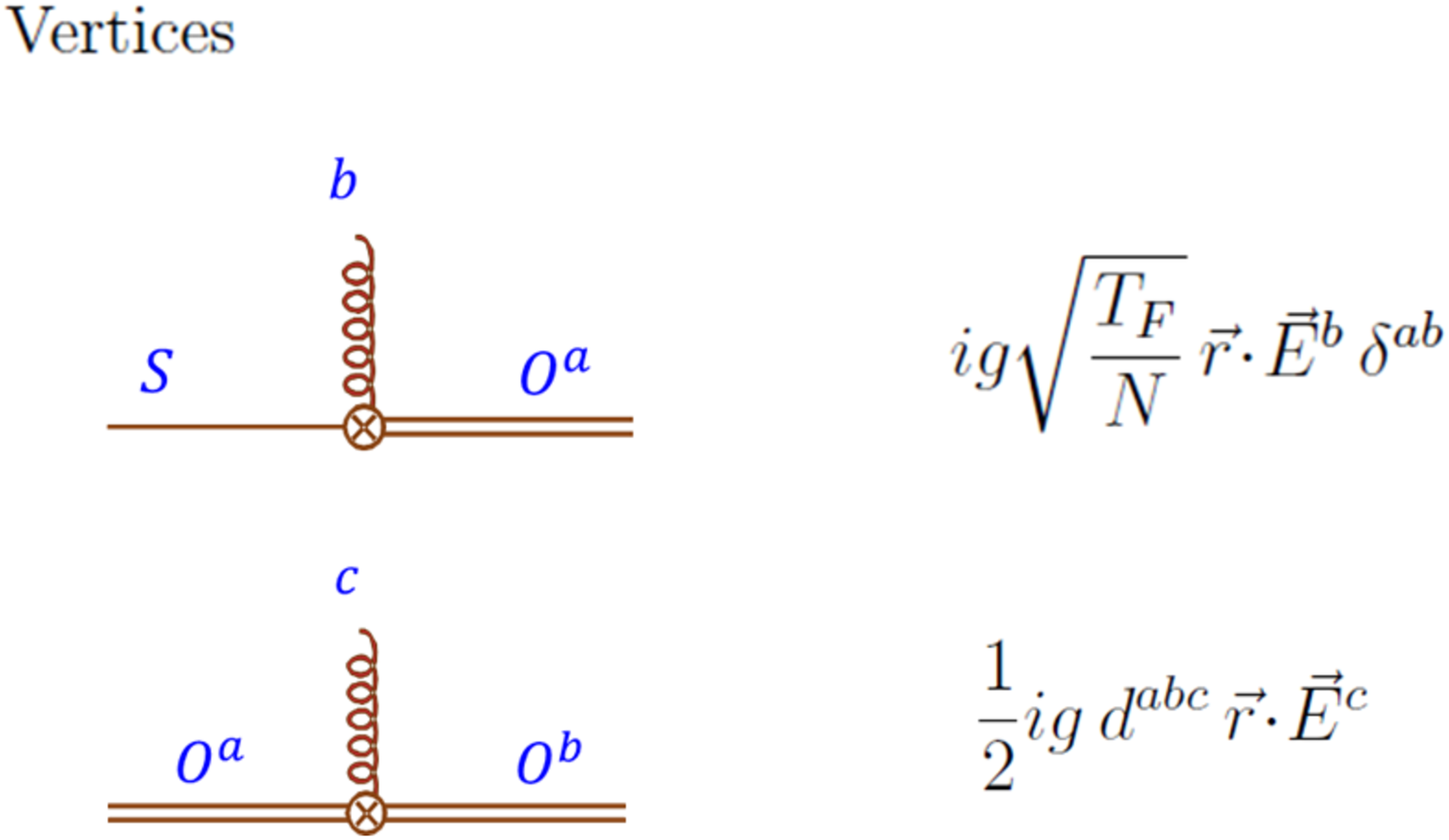}
\vspace{-0.3cm}
\caption{(Color online) Vertices for the $Q\bar{Q}$ singlet to octet transition (upper), and octet to octet transition (lower), due to interaction with an external gluon, with $T_F=1/2$ and $N=N_c=3$. Adapted from~\cite{Sumino:2014qpa}.}
\label{fig_Feynmanrules}
\end{figure}

The color-electric dipole coupling of the $Q\bar{Q}$ system to external soft gluons, being also a core concept of the later developed pNRQCD~\cite{Brambilla:1999xf}, was first derived by Peskin in a seminal operator-product-expansion analysis of how a heavy quark system interacts with external light degrees of freedom~\cite{Peskin:1979va}. A key observation made by Peskin was, to arrive at the dipole coupling, one needs to sum up all possible ways of coupling of an external gluon to the $Q\bar{Q}$ system, in particular including the one peculiar to QCD in which the external gluon couples to the gluon exchanged between the $Q$ and $\bar{Q}$~\cite{Peskin:1979va}. Peskin's perturbative analysis was promoted to the effective Lagrangian level from the perspective of multipole expansion of QCD, which can further be transcribed into a nonrelativistic effective Hamiltonian~\cite{Yan:1980uh}

\begin{align}\label{Heff}
H_{\rm eff}&=H_0 + H_I, \nonumber\\
H_0&=\frac{\vec p^2}{m_Q}+V_1(|\vec r|)+\sum_a\frac{\lambda^a}{2}\frac{\bar{\lambda}^a}{2}V_2(|\vec r|), \nonumber\\
H_I&=V_{SO}+V_{OO}+V_{3g},
\end{align}
where $V_1$ and $V_2$ are the $Q\bar{Q}$ potentials arising from gluon exchange ($\vec r$ being the relative $Q\bar{Q}$ separation) in color singlet and octet configurations, respectively; together with the kinetic energy term, they make up the zeroth order Hamiltonian $H_0$ of the $Q\bar{Q}$ system. The coupling of the $Q\bar{Q}$ system to the external soft gluons $H_I$ consists of also two parts: $V_{SO}$ being the $Q\bar{Q}$ singlet $|S>$ to octet $|O>$ transition vertex (through interacting with an external gluon) that corresponds to a matrix element
\begin{align}\label{V_SO}
<O,a|V_{SO}|S>&=<O,a|\frac{1}{2}g_s\vec r(\frac{\lambda^b}{2}-\frac{\bar{\lambda}^b}{2})\cdot \vec E^b|S>\\ \nonumber
&=\frac{g_s}{\sqrt{2N_c}}\vec E^b\cdot<O|\vec r|S>\delta^{ab},
\end{align}
and $V_{OO}$ associated with the octet $|O>$ to octet transition with the matrix element

\begin{align}\label{V_OO}
<O,a|V_{OO}|O,b>=\frac{ig_s}{2}d^{abc}\vec E^c\cdot <O|\vec r|O>.
\end{align}
These two kinds of vertices are pictorially represented in Fig.~\ref{fig_Feynmanrules}. In Eqs.~(\ref{V_SO}) and (\ref{V_OO}), $g_s$ is the strong coupling constant, $N_c=3$ the number of colors in the fundamental representation of $SU(3)$, and $\lambda^a/2$ and $\bar{\lambda}^a/2$ the color matrices of $Q$ and $\bar{Q}$, respectively; the $a,b,c$ denote the color indices of the $(Q\bar{Q})_8$ octet or the gluons, $d^{abc}=2tr[\lambda^a/2\{\lambda^b/2,\lambda^c/2\}]$ a totally symmetric $SU(3)$ group invariant, and $E^a_i=\partial_iA^a_0-\partial_0A^a_i+g_sf^{abc}A^b_iA^c_0$ the color-electric field of the gluons with color $a$, which reduces to $\vec E^a=-\frac{\partial \vec A^a}{\partial t}$ in the Weyl gauge $A^a_0=0$.

In the interaction Hamiltonian of Eq.~(\ref{Heff}), we add the three-gluon vertex $V_{3g}$ to account for the self-interaction of {\it external} gluons at the order under consideration (four-gluon vertex acts at higher order in $g_s$). In the Weyl gauge adopted here, $V_{3g}$ arises from the energy density $1/2\vec B^a\cdot \vec B^a$ of color-magnetic field $\vec B^{a}(t, \vec x)=\bigtriangledown \times \vec A^a-1/2g_sf^{abc}\vec A^b\times \vec A^c$, such that\cite{Peskin:1995ev}

\begin{align}\label{V3g}
V_{3g}&=(-\frac{g_s}{4})f^{abc}\int d^{3}\vec x[(\bigtriangledown\times \vec A^a)\cdot(\vec A^b\times \vec A^c) \nonumber \\ & +(\vec A^b\times \vec A^c)\cdot(\bigtriangledown\times \vec A^a)],
\end{align}
where $f^{abc}$ is the antisymmetric structure constant of the $SU(3)$ color group. The transverse gluon field can be quantized
\begin{equation}\label{Aa}
\vec A^a(t,\vec x)=\sum_{\vec k,\lambda}N_{\vec k}\vec \epsilon_{\vec k\lambda}[a_{\vec k\lambda}^a e^{i\vec k\cdot\vec x-i\omega_{\vec k}t}+h.c.],
\end{equation}
where $\vec k$ is the gluon momentum, $\omega_{\vec k}$ the energy, and $\epsilon_{\vec k \lambda=1,2}$ the two physical polarization vectors. With normalization constant $N_{\vec k}=\sqrt{\frac{\hbar c^2}{2V\omega_{\vec k}}}$ ($V$ being the spatial volume) in the rationalized Gauss unit as used here, the creation and annihilation operators of gluons in Eq.~(\ref{Aa}) satisfy the commutation relation $[a_{\vec k\lambda}^a, a_{\vec k\,'\lambda\,'}^{b\dagger}]=\delta_{\vec k\vec k\,'}\delta_{\lambda\lambda\,'}\delta^{ab}$. While the derivation of the LO (gluo-dissociation) cross section involves only the singlet to octet transition vertex $V_{SO}$~\cite{Chen:2017jje}, in deriving the NLO cross section, the octet to octet transition vertex $V_{OO}$ and the three-gluon vertex $V_{3g}$  will be also invoked.

%%%%%%%%%%%%%%%%%%%%%%%%%%%%%%%%%%%%%%%%%%%%%%%%%%%%%%%%%%%%%%%%
\subsection{Derivation of the NLO cross section}
\label{ssec_deriveJpsiNLOcrosssecion}
%%%%%%%%%%%%%%%%%%%%%%%%%%%%%%%%%%%%%%%%%%%%%%%%%%%%%%%%%%%%%%%%

We now use the effective Hamiltonian specified above to derive the NLO dissociation cross section of the $1S$ charmonium state by an external gluon, {\it i. e.}, the process of $g+J/\psi\rightarrow g+ c+\bar{c}$. We work in the framework of the second order quantum-mechanical perturbation and use the natural units $\hbar=c=1$ throughout.

We start with the standard expression of the transition amplitude in second order perturbation theory
\begin{equation}\label{2ndTfi}
T_{fi}=\sum_m\frac{<f|H_I|m><m|H_I|i>}{E_i-E_m+i\epsilon},
\end{equation}
(an overall phase factor $e^{i(E_i-E_f)t}$ that does not affect the calculation of cross section is neglected) where the initial state involves a bound state $J/\psi$ plus an incident gluon of momentum $\vec k$, polarization $\lambda$ and color $a$: $|i>=|J/\psi,g(\vec k,\lambda,a)>$, and the final state involves an unbound octet $(c\bar{c})_8$ of internal relative (between $c$ and $\bar{c}$) momentum $\vec p$ and color $b$, plus an outgoing gluon of momentum $\vec \kappa$, polarization $\sigma$ and color $c$: $|f>=|(c\bar{c})_8(\vec p,b),g(\vec \kappa,\sigma,c)>$. Note that we work in the rest frame of the $J/\psi$ and in the same spirit of calculating the LO (gluo-dissociation) cross section, neglect the three-momentum transferred by the incident gluon to the $J/\psi$. The latter approximation is justified by the fact that the mass of the $Q\bar{Q}$ bound states ({\it e.g.}, the mass of $J/\psi$ is 3.097\,GeV, and of $\Upsilon(1S)$ is 9.460\,GeV) is much larger than the typical momentum of thermal gluons in QGP which is of the order of temperature. As a result, the center-of-mass momentum of the final state unbound octet $(c\bar{c})_8$ is neglected; that is, the rest frame of $J/\psi$ is approximately also the rest frame of the $(c\bar{c})_8$, and then one needs only to deal with the internal relative momentum $\vec p$ of the $(c\bar{c})_8$. Such an approximation is apparently better for the more massive bottomonium states.

Eq.~(\ref{2ndTfi}) involves a summation over all possible intermediate states $|m>$. The transition $|i>\rightarrow|f>$ considered here can take place upon either the successive action of $V_{SO}$ and $V_{OO}$ (corresponding to $s$- or $u$-channel reaction), or of $V_{SO}$ and $V_{3g}$ (corresponding to $t$-channel reaction). As required by particle number conservation, the intermediate states allowed in the former case are  $(a).~|m>=|(c\bar{c})_8(\vec q,d)>$ and $(b).~|m>=|(c\bar{c})_8(\vec q,d),g(\vec k_1,\lambda_1,d_1),g(\vec k_2,\lambda_2,d_2)>$; and intermediate states allowed in the latter case are $(c).~|m>=|(c\bar{c})_8(\vec q,d),g(\vec k_1,\lambda_1,d_1),g(\vec k_2,\lambda_2,d_2)>$ and $(d).~|m>=|J/\psi,g(\vec k_1,\lambda_1,d_1),g(\vec k_2,\lambda_2,d_2)>$. It turns out there's no interference between the amplitude of $(a)+(b)$ and that of $(c)+(d)$. So in the following, we discuss the computations of these two cases separately.

With the first kind of intermediate states $(a).~|m>=|(c\bar{c})_8(\vec q,d)>$, the matrix element $<m|V_{SO}|i>$ in the numerator of Eq.~(\ref{2ndTfi}) can be computed by using Eqs.~(\ref{V_SO}) and (\ref{Aa})
\begin{align}
&<m|V_{SO}|i>=\delta^{ad}\frac{ig_s}{2}\sqrt{\frac{\omega_{\vec k}}{3V}}<(c\bar{c})_8(\vec q)|\vec r\cdot\vec \epsilon_{\vec k\lambda}|J/\psi>\nonumber \\
&=\delta^{ad}\frac{-g_s\sqrt{\omega_{\vec k}}}{2\sqrt{3}V}\vec \epsilon_{\vec k\lambda}\cdot\nabla_{\vec q}\int d^3r e^{-i\vec q\cdot\vec r}R_{10}(r)Y_{00}(\theta,\phi)\nonumber\\
&=\delta^{ad}\frac{(-g_s)}{V}\sqrt{\frac{\pi\omega_{\vec k}}{3}}~\vec \epsilon_{\vec k\lambda}\cdot\frac{\vec q}{q}\int r^3 drj_1(qr)R_{10}(r),
\end{align}
where $R_{10}(r)$ is the normalized radial wave function of the $1S$ bound state $J/\psi$. We have neglected the interaction between $c$ and $\bar{c}$ in the octet state and therefore it is represented by a plane wave. Upon a spherical wave expansion $e^{-i\vec p\cdot\vec r}=4\pi\sum_l\sum_m(-i)^lj_l(pr)Y_{lm}(\theta,\phi)Y_{lm}(\theta\,',\phi\,')$ (primed angels for $\vec p$, and unprimed for $\vec r$) and using the orthogonality relation for the spherical harmonics $\int\int {\rm sin}\theta d\theta d\phi Y_{lm}^*(\theta,\phi)Y_{l\,'m\,'}(\theta,\phi)=\delta_{ll\,'}\delta_{mm\,'}$, and integrating out the coordinate angles, one is left with only the spherical Bessel function of the first order $j_1(qr)$.

In a similar way, the other matrix element in the numerator of Eq.~(\ref{2ndTfi}) can be computed by using Eqs.~(\ref{V_OO}) and (\ref{Aa})
\begin{align}
&<f|V_{OO}|m>=\frac{g_s}{2}\sqrt{\frac{\omega_{\vec \kappa}}{2V}}\delta^{cf}d^{bdf}\vec \epsilon_{\vec \kappa\sigma}\cdot<(c\bar{c})_8(\vec p)|\vec r|(c\bar{c})_8(\vec q)>\nonumber \\
&=\frac{(-ig_s)}{2V}(2\pi)^3d^{bdc}\sqrt{\frac{\omega_{\vec \kappa}}{2V}}\epsilon_{\vec \kappa\sigma}\cdot\nabla_{\vec q}\delta^3(\vec q-\vec p),
\end{align}
which involves no bound state wave function. Plugging these two matrix elements into Eq.~(\ref{2ndTfi}) and performing the summation over the momentum $\vec q$ and color $d$ of the intermediate states $|m>=|(c\bar{c})_8(\vec q,d)>$, one obtains the first contribution to the second order transition amplitude
\begin{align}\label{2ndTfi1}
T_{fi}^{(a)}&=\frac{(-ig_s^2)}{2V^2}(2\pi)^3\sqrt{\frac{\pi\omega_{\vec k}\omega_{\vec \kappa}}{6V}}\sum_d\delta^{ad}d^{bdc}\frac{V}{(2\pi)^3}\int d^3\vec q\vec \epsilon_{\vec k\lambda}\cdot \frac{\vec q}{q}\nonumber \\ &\times\int r^3dr j_1(qr)R_{10}(r)\frac{\vec \epsilon_{\vec\kappa\sigma}\cdot\nabla_{\vec q}\delta^3(\vec q-\vec p)}{-\epsilon_B+\omega_{\vec k}-\frac{\vec q^2}{m_Q}+i\epsilon} \nonumber \\
&=d^{bac}\frac{ig_s^2}{2V}\sqrt{\frac{\pi\omega_{\vec k}\omega_{\vec\kappa}}{6V}}\vec\epsilon_{\vec k\lambda}\cdot[A(p,k)\vec\epsilon_{\vec \kappa\sigma}+(\vec\epsilon_{\vec\kappa\sigma}\cdot\vec p)\frac{\vec p}{p^2}B(p,k)],
\end{align}
where
\begin{align}\label{ABpk}
A(p,k)=&\frac{\int r^3 dr j_1(pr)R_{10}(r)}{p(-\epsilon_B+\omega_{\vec k}-\frac{\vec p^2}{m_Q}+i\epsilon)},\nonumber \\
B(p,k)=&\frac{\frac{2p}{m_Q}\int r^3 dr j_1(pr)R_{10}(r)}{(-\epsilon_B+\omega_{\vec k}-\frac{\vec p^2}{m_Q}+i\epsilon)^2}-\frac{\int r^4 dr j_2(pr)R_{10}(r)}{(-\epsilon_B+\omega_{\vec k}-\frac{\vec p^2}{m_Q}+i\epsilon)}.
\end{align}
In Eq.~(\ref{2ndTfi1}), an integration by part with respect to $\int d^3\vec q$ was performed to integrate out the $\delta^3(\vec q-\vec p)$ and arrive at the last line. All energies (including the binding energy $\epsilon_B$ of the bound states) are measured relative to the threshold of the $c\bar{c}$ pair. We note that the appearance of the spherical Bessel function of the second order $j_2(pr)$ here is simply ascribed to the selection rule $\Delta l=1$ of the color-electric dipole transition that now acts twice from the initial $S$-wave ($l=0$) state.

For the second intermediate states $(b).~|m>=|(c\bar{c})_8(\vec q,d),g(\vec k_1,\lambda_1,d_1),g(\vec k_2,\lambda_2,d_2)>$, similar but a bit more lengthy (to handle the two-gluon state with creation and annihilation operators, and also summation over intermediate states now reads $\sum_m=\frac{V}{(2\pi)^3}\int d^3\vec q\sum_d\sum_{\vec k_1\lambda_1d_1}\sum_{\vec k_2\lambda_2d_2}$) manipulations lead to transition amplitude
\begin{align}\label{2ndTfi2}
&T_{fi}^{(b)}=-d^{bac}\frac{ig_s^2}{2V}\sqrt{\frac{\pi\omega_{\vec k}\omega_{\vec \kappa}}{6V}}\nonumber \\
&\times\vec\epsilon_{\vec k\lambda}\cdot[C(p,\kappa)\vec\epsilon_{\vec \kappa\sigma}+(\vec\epsilon_{\vec \kappa\sigma}\cdot\vec p)\frac{\vec p}{p^2}D(p,\kappa)],
\end{align}
where the property of $d^{abc}$ being totally symmetric has been exploited to simplify the algebras, and
\begin{align}\label{CDpkappa}
C(p,\kappa)&=\frac{\int r^3 dr j_1(pr)R_{10}(r)}{p(-\epsilon_B-\omega_{\vec \kappa}-\frac{\vec p^2}{m_Q}+i\epsilon)},\nonumber \\
D(p,\kappa)&=\frac{\frac{2p}{m_Q}\int r^3 dr j_1(pr)R_{10}(r)}{(-\epsilon_B-\omega_{\vec \kappa}-\frac{\vec p^2}{m_Q}+i\epsilon)^2}-\frac{\int r^4 dr j_2(pr)R_{10}(r)}{(-\epsilon_B-\omega_{\vec \kappa}-\frac{\vec p^2}{m_Q}+i\epsilon)}.
\end{align}
So both $T_{fi}^{(a)}$, $T_{fi}^{(b)}$ $\varpropto d^{abc}$.

Now we move to the transition due to successive action of $V_{SO}$ and $V_{3g}$. For the third intermediate state, $(c).~|m>=|(c\bar{c})_8(\vec q,d),g(\vec k_1,\lambda_1,d_1),g(\vec k_2,\lambda_2,d_2)>$,
\begin{align}
<m|V_{SO}|i>&=\frac{g_s}{V}\sqrt{\frac{\pi\omega_{\vec k_1}}{3}}\delta_{\vec k_2\vec k}\delta_{\lambda_2\lambda}\delta^{d_2a}\delta^{d_1d}\vec \epsilon_{\vec k_1\lambda_1}\cdot\frac{\vec q}{q}\nonumber \\
&\times\int r^3 drj_1(qr)R_{10}(r),
\end{align}
and
\begin{align}
&<f|V_{3g}|m>=\delta^{bd}\frac{(2\pi)^3}{V}\delta^3(\vec p-\vec q)\nonumber\\
&\times <g(\vec \kappa,\sigma,c)|V_{3g}|g(\vec k_1,\lambda_1,d_1),g(\vec k_2,\lambda_2,d_2)>\nonumber\\
&=(-\frac{g_s}{2})\frac{(2\pi)^3}{V}\delta^3(\vec p-\vec q)\sqrt{\frac{1}{2V\omega_{\vec k_1}\omega_{\vec k_2}\omega_{\vec \kappa}}}\delta^{bd}f^{d_1d_2c}\delta_{\vec k_1+\vec k_2,\vec\kappa}\nonumber\\
&\times [(\vec \epsilon_{\vec k_1\lambda_1}\times i\vec k_1)\cdot (\vec\epsilon_{\vec k_2\lambda_2}\times\vec\epsilon_{\vec\kappa\sigma})-(\vec\epsilon_{\vec k_2\lambda_2}\times i\vec k_2)\cdot (\vec\epsilon_{\vec k_1\lambda_1}\times\vec\epsilon_{\vec\kappa\sigma})\nonumber\\
&-(\vec\epsilon_{\vec\kappa\sigma}\times i\vec \kappa)\cdot (\vec\epsilon_{\vec k_1\lambda_1}\times \vec\epsilon_{\vec k_2\lambda_2})].
\end{align}
Combining $<m|V_{SO}|i>$ and $<f|V_{3g}|m>$ above and performing the summation over intermediate states, one arrives at the transition amplitude
\begin{align}\label{2ndTfi1c}
&T_{fi}^{(c)}=f^{abc}\frac{(-ig_s^2)}{V}\sqrt{\frac{\pi}{6V \omega_{\vec k}\omega_{\vec \kappa}}}\{(\vec \epsilon_{\vec \kappa \sigma} \cdot \vec p)(\vec \epsilon_{\vec k \lambda} \cdot \vec \kappa)\nonumber\\ &+(\vec \epsilon_{\vec \kappa \sigma} \cdot \vec k)(\vec \epsilon_{\vec k \lambda} \cdot \vec p)
-(\vec \epsilon_{\vec k \lambda} \cdot \vec \epsilon_{\vec \kappa \sigma})\frac{(\vec k^{2}-\vec k \cdot \vec \kappa)\vec \kappa \cdot \vec p+(\vec \kappa^{2}-\vec k \cdot \vec \kappa)\vec k \cdot \vec p}{(\vec k-\vec \kappa)^{2}}\}\nonumber\\
&\times\frac{1}{p}\int r^3dr j_1(pr)R_{10}(r)\frac{1}{-\epsilon_{B}^{J/\psi}-\frac{\vec p^2}{m_Q}-\omega(\vec k-\vec \kappa)+i\epsilon}.
\end{align}

Similarly, for the fourth intermediate state $(d).~|m>=|J/\psi,g(\vec k_1,\lambda_1,d_1),g(\vec k_2,\lambda_2,d_2)>$, the transition amplitude
\begin{align}\label{2ndTfi1d}
&T_{fi}^{(d)}=f^{abc}\frac{ig_s^2}{V}\sqrt{\frac{\pi}{6V \omega_{\vec k}\omega_{\vec \kappa}}}\{(\vec \epsilon_{\vec \kappa \sigma} \cdot \vec p)(\vec \epsilon_{\vec k \lambda} \cdot \vec \kappa)+(\vec \epsilon_{\vec \kappa \sigma} \cdot \vec k)(\vec \epsilon_{\vec k \lambda} \cdot \vec p) \nonumber\\
&-(\vec \epsilon_{\vec k \lambda} \cdot \vec \epsilon_{\vec \kappa \sigma})\frac{(\vec k^{2}-\vec k \cdot \vec \kappa)\vec \kappa \cdot \vec p+(\vec \kappa^{2}-\vec k \cdot \vec \kappa)\vec k \cdot \vec p}{(\vec k-\vec \kappa)^{2}}\}\nonumber\\
&\times\frac{1}{p}\int r^3dr j_1(pr)R_{10}(r)\frac{1}{\omega(\vec k)-\omega(\vec \kappa)-\omega(\vec k-\vec \kappa)+i\epsilon}.
\end{align}
Therefore both $T_{fi}^{(c)}$, $T_{fi}^{(d)}$ $\varpropto f^{abc}$.

We note that $d^{abc}f^{abc}=0$, so that there's no interference between the amplitude $T_{fi}^{(a)}+T_{fi}^{(b)}$ and the amplitude $T_{fi}^{(c)}+T_{fi}^{(d)}$; that is, they have independent cross sections. For $T_{fi}^{(a)}+T_{fi}^{(b)}$, the corresponding second order transition rate
is given by
\begin{align}
\Gamma_{i\rightarrow f}^{(a+b)}=\frac{2\pi}{\hbar}\sum_f|T_{fi}^{(a)} + T_{fi}^{(b)}|^2\delta(E_i-E_f),
\end{align}
which, upon dividing by the incident flux $v_{rel}/V$ ($v_{rel}$ being the relative velocity between the incident gluon and the target $J/\psi$) and averaging (summing) over initial (final) state degeneracies, is converted into the NLO cross section
\begin{align}
&\sigma^{(a+b)}(E_g)=2\pi V\frac{V}{(2\pi)^3}\int d^3\vec p\sum_b\frac{V}{(2\pi)^3}\int d^3\vec\kappa\sum_\sigma\sum_c\frac{1}{4\pi}\int d\Omega_{\vec k}\nonumber \\
&\times\frac{1}{2}\sum_{\lambda}\frac{1}{8}\sum_a|T_{fi}^{(a)} + T_{fi}^{(b)}|^2\delta(-\epsilon_B+\omega_{\vec k}-\frac{\vec p^2}{m_Q}-\omega_{\vec\kappa}),
\end{align}
where $E_g=\omega_{\vec k}$ is the incident gluon energy in the rest frame of $J/\psi$. The average (summation) over initial (final) state degeneracies can be readily performed by using the identity
\begin{equation}\label{polarizationaverage1}
\frac{1}{4\pi} \int d\Omega_{\vec k}\frac{1}{2}\sum_{\lambda=1,2}|\vec\epsilon_{\vec k\lambda}\cdot\vec \rho|^2=\frac{1}{3}|\vec \rho|^2
\end{equation}
twice ($\vec\rho$ being an arbitrary vector independent of $\Omega_{\vec k}$), the first for the polarization vector of the incident gluon $\vec\epsilon_{\vec k\lambda}$, and the second for that of the outgoing gluon $\vec\epsilon_{\vec\kappa\sigma}$ in the final state. Further using the identity $\sum_{abc}d^{bac}d^{bac}=40/3$, the cross section can be finally computed from
\begin{align}\label{NLOcrosssectionab}
&\sigma^{(a+b)}(E_g)=\frac{5}{216\pi^2}g_s^4E_g\int p^2 dp\int \kappa^2d\kappa\omega_{\vec\kappa}\nonumber \\
&\times \{\cdots\}\delta(-\epsilon_B+E_g-\frac{\vec p^2}{m_Q}-\omega_{\vec\kappa}),
\end{align}
~~~~~~where
\begin{align}\label{sigmakernal}
\{\cdots\}&=[A^2(p,k)+\frac{1}{3}B^2(p,k)+\frac{2}{3}A(p,k)B(p,k)]\nonumber \\&+[C^2(p,\kappa)+\frac{1}{3}D^2(p,\kappa)+\frac{2}{3}C(p,\kappa)D(p,\kappa)]\nonumber \\
&-2[A(p,k)C(p,\kappa)+\frac{1}{3}(A(p,k)D(p,\kappa)\nonumber \\
&+B(p,k)C(p,\kappa)+B(p,k)D(p,\kappa))].
\end{align}
Note that in Eqs.~(\ref{ABpk}) and (\ref{CDpkappa}), upon using the  energy-conserving $\delta$-function, the denominators $\varpropto \omega_{\vec \kappa}, \omega_{\vec k}$, respectively, so that the zero point is not reached and thus the $i\epsilon$ factor can be dropped. To see how the integrations over final state momenta in Eq.~(\ref{NLOcrosssectionab}) work out with $\delta$-function, we define $p_c=\sqrt{(E_g-\epsilon_B-m_g)m_Q}$, with $m_g$ being an effective gluon mass and $\omega_{\vec\kappa}=\sqrt{\vec\kappa^2+m_g^2}$. Apparently, if $p>p_c$, one has $-\epsilon_B+E_g-\frac{\vec p^2}{m_Q}=\frac{\vec p_c^2}{m_Q}+m_g-\frac{p^2}{m_Q}<m_g\leq\omega_{\vec\kappa}$, so that the zero point of the argument of the $\delta$-function can never be reached for any $\kappa>0$, and therefore the corresponding integrand vanishes; that is to say, $p_c$ serves as a cut-off for the integration over $p$. On the other hand, for $p<p_c$, the $\delta$-function can be integrated out via $\int d\kappa$, which, together with $\int_0^{p_c}dp$, yields a finite result. However, if the gluon is massless, $A(p,k)$ and $B(p,k)$ of Eq.~(\ref{ABpk}) would be $\varpropto 1/\omega_{\vec \kappa}=1/\kappa$, causing an infrared divergence (from $\kappa\rightarrow 0$) that manifests itself in the integration $\int_0^{p_c}dp$ as $p\rightarrow p_c$. In contrast, no divergence would occur if the initial state incident gluon is massless (the same is true for LO dissociation cross section~\cite{Chen:2017jje}), as the denominator of $C(p,\kappa)$ and $D(p,\kappa)$ (see Eq.~(\ref{CDpkappa})) $\varpropto \omega_{\vec k}=k$ must be larger than $\epsilon_B$ in order to overcome the bound state binding energy.

Similarly, for $T_{fi}^{(c)}+T_{fi}^{(d)}$, the cross section is derived, which reads
\begin{align}\label{NLOcrosssectioncd}
&\sigma^{(c+d)}(E_g)=\frac{g_s^4}{8\pi^4}\frac{1}{E_g}\int dp\int d\kappa\frac{\kappa^2}{\omega(\vec \kappa)}\int d\Omega_{\vec \kappa}\int d\Omega_{\vec k}\nonumber\\
&[\int r^3 drj_{1}(pr)R_{10}(r)]^2\times g(\vec \kappa,\vec p, \vec k)\nonumber\\
&\times[\frac{\omega(\vec \kappa)-\omega(\vec k)}{\omega^2(\vec k-\vec \kappa)-(\omega(\vec k)-\omega(\vec \kappa))^2}]^2\delta(-\epsilon_B+E_g-\frac{\vec p^2}{m_Q}-\omega_{\vec\kappa}),
\end{align}
where $g(\vec \kappa,\vec p, \vec k)$ is a polynomial function of these three momenta, that arises from averaging over polarization vectors $\vec \epsilon_{\vec \kappa\sigma}$ and $\vec \epsilon_{\vec k\lambda}$ with
\begin{align}\label{polarizationaverage2}
\sum_{\sigma}\epsilon_{\vec \kappa\sigma}^{(i)}\epsilon_{\vec \kappa\sigma}^{(j)}=\delta^{ij}-\frac{\kappa^i\kappa^i}{\vec \kappa^2}.
\end{align}
Here because of explicit angular dependence, Eq.~(\ref{polarizationaverage1}) ceases to function in favor of Eq.~(\ref{polarizationaverage2}). The $\delta$-function in Eq.~(\ref{NLOcrosssectioncd}) is handled in the same way as discussed above. For massless {\it and} collinear ({\it i.e.}, $\vec k$ $//$ $\vec \kappa$) gluons, the denominator of the term in the square bracket of Eq.~(\ref{NLOcrosssectioncd}) is vanishing; this soft-collinear divergence will be regularized by finite thermal gluon masses.

\begin{figure} [!t]
\includegraphics[width=1.05\columnwidth]{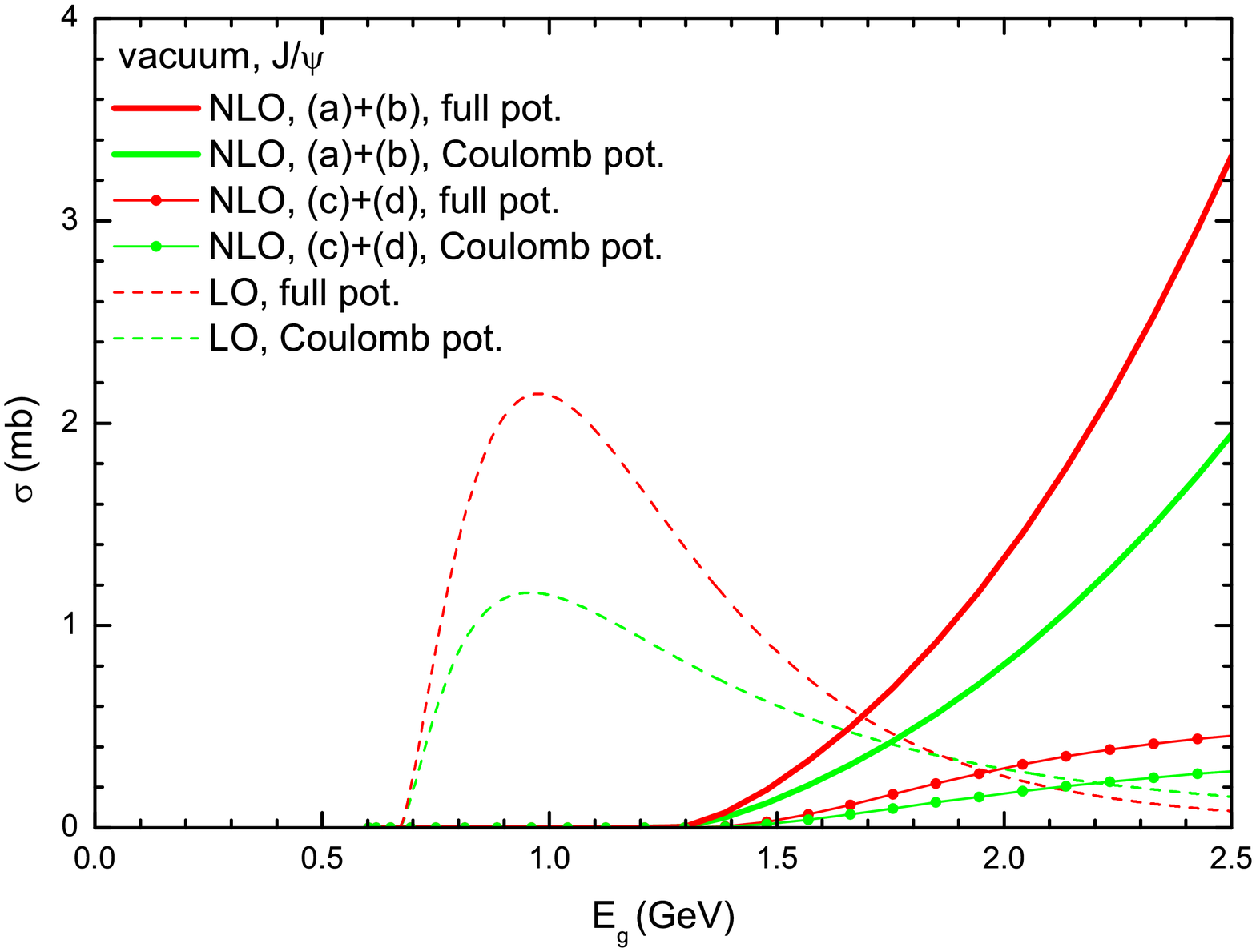}
\vspace{-0.3cm}
\includegraphics[width=1.05\columnwidth]{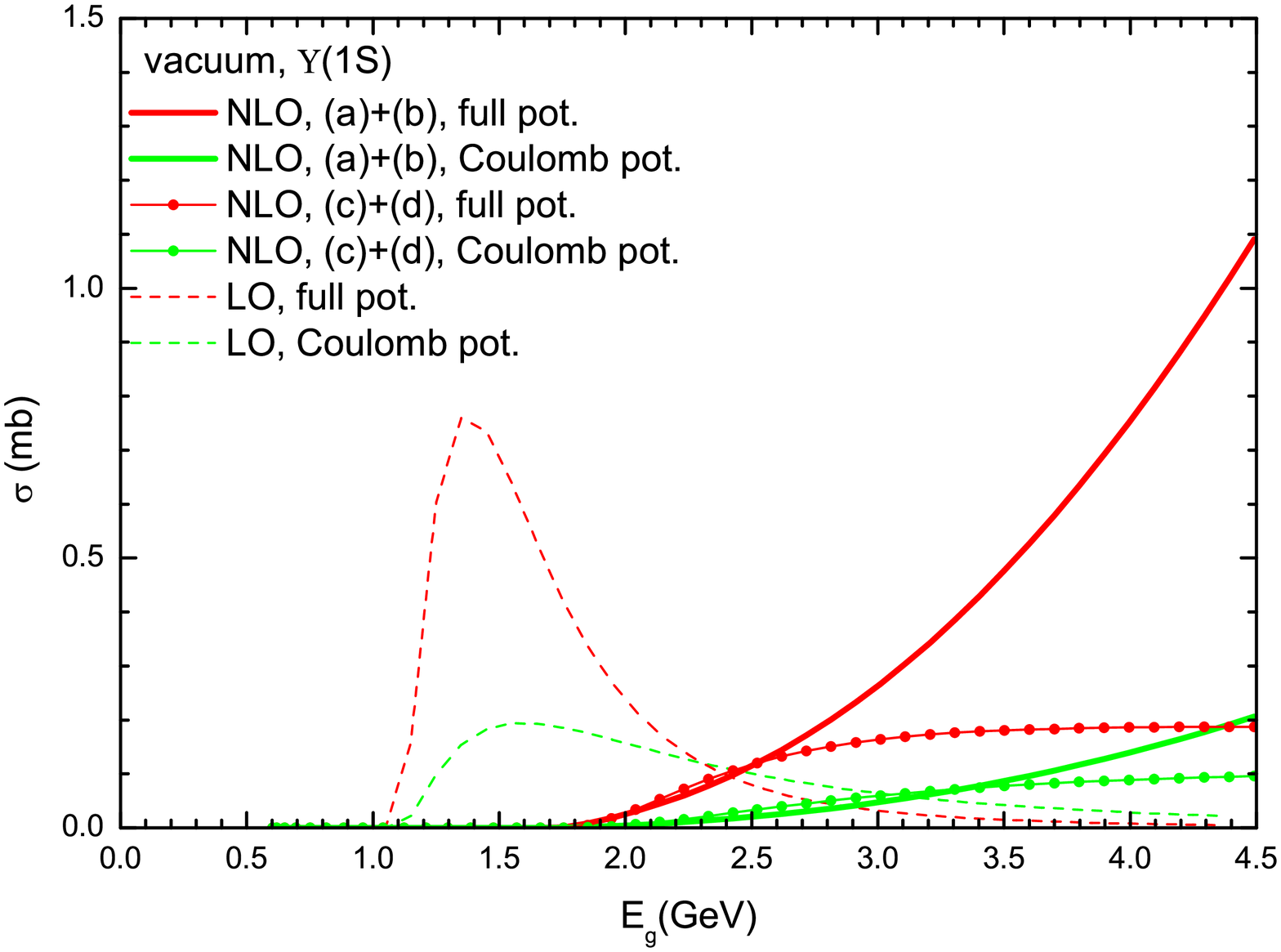}
\vspace{-0.3cm}
\caption{(Color online) NLO cross sections (from amplitudes $(a)+(b)$ and $(c)+(d)$, respectively) of $J/\psi$ (upper panel), and $\Upsilon(1S)$ (lower panel) with a gluon of effective mass $m_g=600$\,MeV, in comparison with the corresponding LO (gluo-dissociation) cross section. Vacuum bound state wave functions with Coulomb potential and full Cornell potential are used and compared (see text for more details).}
\label{fig_vacuumcrosssections}
\end{figure}

To sum up, we've demonstrated the derivation of the NLO dissociation cross section for the $1S$ state $J/\psi$. The same has been also derived for the $P$-wave state $\chi_c$, which is a bit more complicated because of dependence of the pertinent wave function on the azimuthal angle. The NLO cross section thus derived is divergent (infrared and soft-collinear divergence) if the gluons are massless. Nevertheless, to make a comparison with the LO (gluo-dissociation) cross section that is well defined even in vacuum, we assume an effective mass of the typical constituent value $m_g=600$\,MeV for the gluon and evaluate the NLO cross section with vacuum bound state wave function. The results are displayed in Fig.~\ref{fig_vacuumcrosssections} for $J/\psi$ and $\Upsilon(1S)$, where solutions of the bound state wave functions with Coulomb or full Cornell potential from Schr\"{o}dinger equation (see the next section for more details) are also compared. Unlike the LO cross section falling off toward higher gluon energies, the NLO cross section from amplitude $(a)+(b)$ (Eq.~(\ref{NLOcrosssectionab})), with the beginning point delayed (the incident gluon now needs to overcome the binding energy plus the final state gluon mass in order to break up the bound state), increases monotonously with the incident gluon energy, while the NLO cross section from amplitude $(c)+(d)$ (Eq.~(\ref{NLOcrosssectioncd})) quickly levels off.

%%%%%%%%%%%%%%%%%%%%%%%%%%%%%%%%%%%%%%%%%%%%%%%%%%%%%%%%%%%%%%%%
\section{NLO Dissociation of Various Heavy Quarkonia in an In-medium Potential Model}
\label{sec_in-medium_model}
%%%%%%%%%%%%%%%%%%%%%%%%%%%%%%%%%%%%%%%%%%%%%%%%%%%%%%%%%%%%%%%%

%%%%%%%%%%%%%%%%%%%%%%%%%%%%%%%%%%%%%%%%%%%%%%%%%%%%%%%%%%%%%%%%
\subsection{In-medium NLO cross sections}
\label{ssec_disscrosssections}
%%%%%%%%%%%%%%%%%%%%%%%%%%%%%%%%%%%%%%%%%%%%%%%%%%%%%%%%%%%%%%%%
We now turn to the calculation of in-medium NLO dissociation cross section of heavy quarkonia by thermal gluons. In QGP, the gluons acquire a temperature dependent thermal mass, which is taken to be $m_g(T)=\sqrt{3/4}g_sT$ with fixed $g_s=2.3$ for $N_f=3$ active light flavors and $N_c=3$ colors~\cite{Riek:2010fk}, rendering the NLO cross section finite. For the in-medium bound state radial wave functions $R_{nl}(r)$ that enter the evaluation of the cross sections, we solve the radial Schr\"{o}dinger equation of the $Q\bar{Q}$ system
\begin{align}
&\frac{1}{r^2}\frac{d}{dr}(r^2\frac{dR_{nl}}{dr})+\nonumber \\
&[m_Q(E_{n,l}-2m_Q-V(r,T))-\frac{l(l+1)}{r^2}]R_{nl}(r)=0,
\end{align}
with a temperature dependent potential parameterized in~\cite{Karsch:1987pv}
\begin{equation}\label{fullpotential}
V(r,T)=-\frac{\alpha}{r}e^{-m_D(T)r}+\frac{\sigma}{m_D(T)}(1-e^{-m_D(T)r}),
\end{equation}
which is a modification of the vacuum Cornell potential $V(r,0)=-\alpha/r+\sigma r$ by color screening~\cite{Karsch:1987pv}, with the Debye screening mass $m_D/T=-4.058 + 6.32\cdot(T/T_c-0.885 )^{0.1035}$ ($T_c=172.5$\,MeV) fitted to lattice data from~\cite{Burnier:2015tda,Chen:2017jje}. With coupling strength $\alpha=4/3\alpha_s=0.471$ of the Coulomb part, the string tension $\sigma=0.192\,\rm GeV^2$ of the confining part, and heavy quark masses $m_c=1.320$\,GeV, $m_b=4.746$\,GeV, the vacuum masses of charmonia and bottomonia below threshold are well reproduced~\cite{Karsch:1987pv,Chen:2017jje}. The temperature-dependent binding energy of the bound state is then obtained via $\epsilon_B(T)=2m_Q+\sigma/m_D(T)-E_{n,l}(T)$, whose zero point defines the dissociation temperature of the bound state under consideration~\cite{Karsch:1987pv,Chen:2017jje}.

\begin{figure} [!t]
\includegraphics[width=1.05\columnwidth]{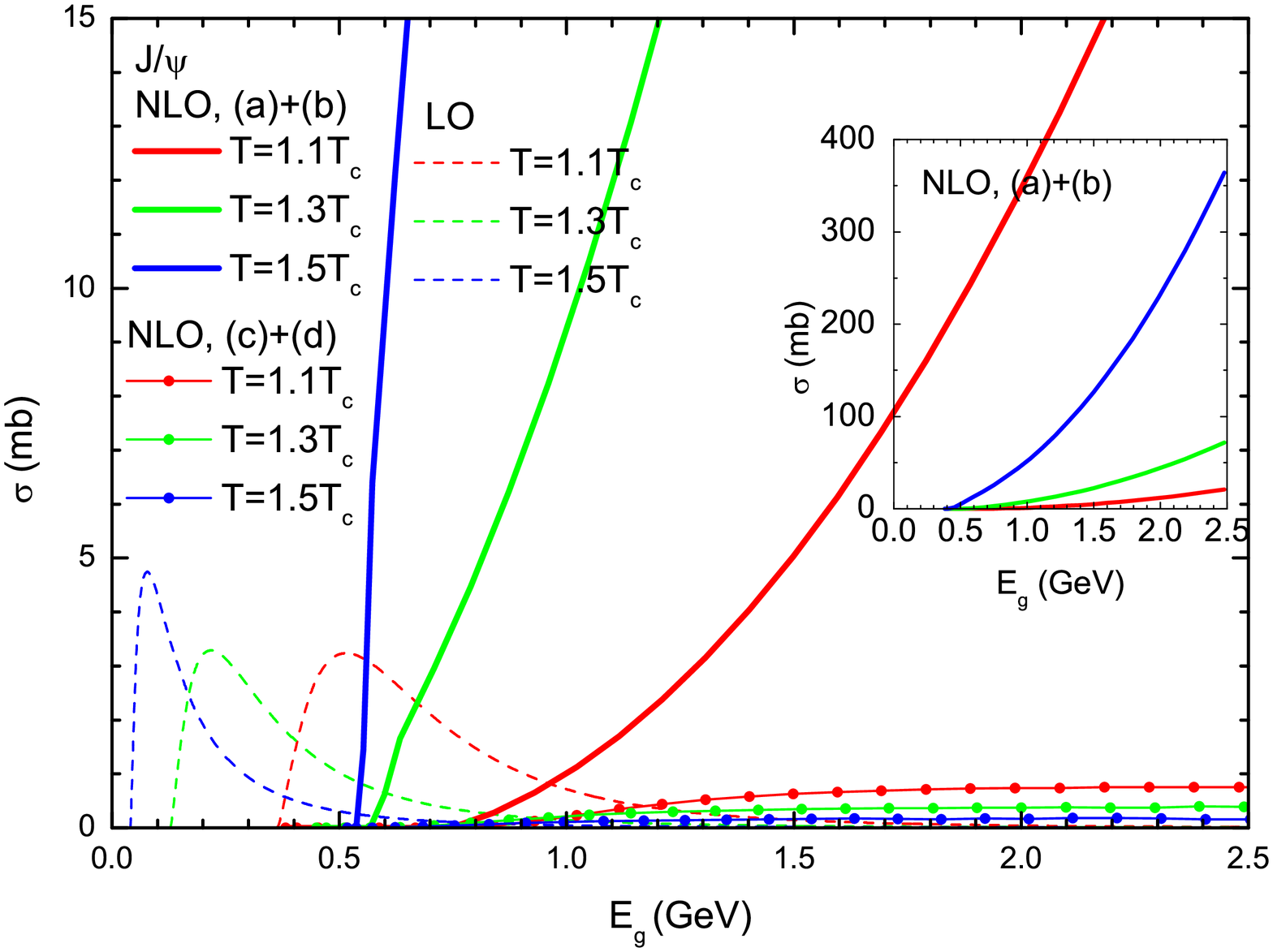}
\vspace{-0.3cm}
\includegraphics[width=1.05\columnwidth]{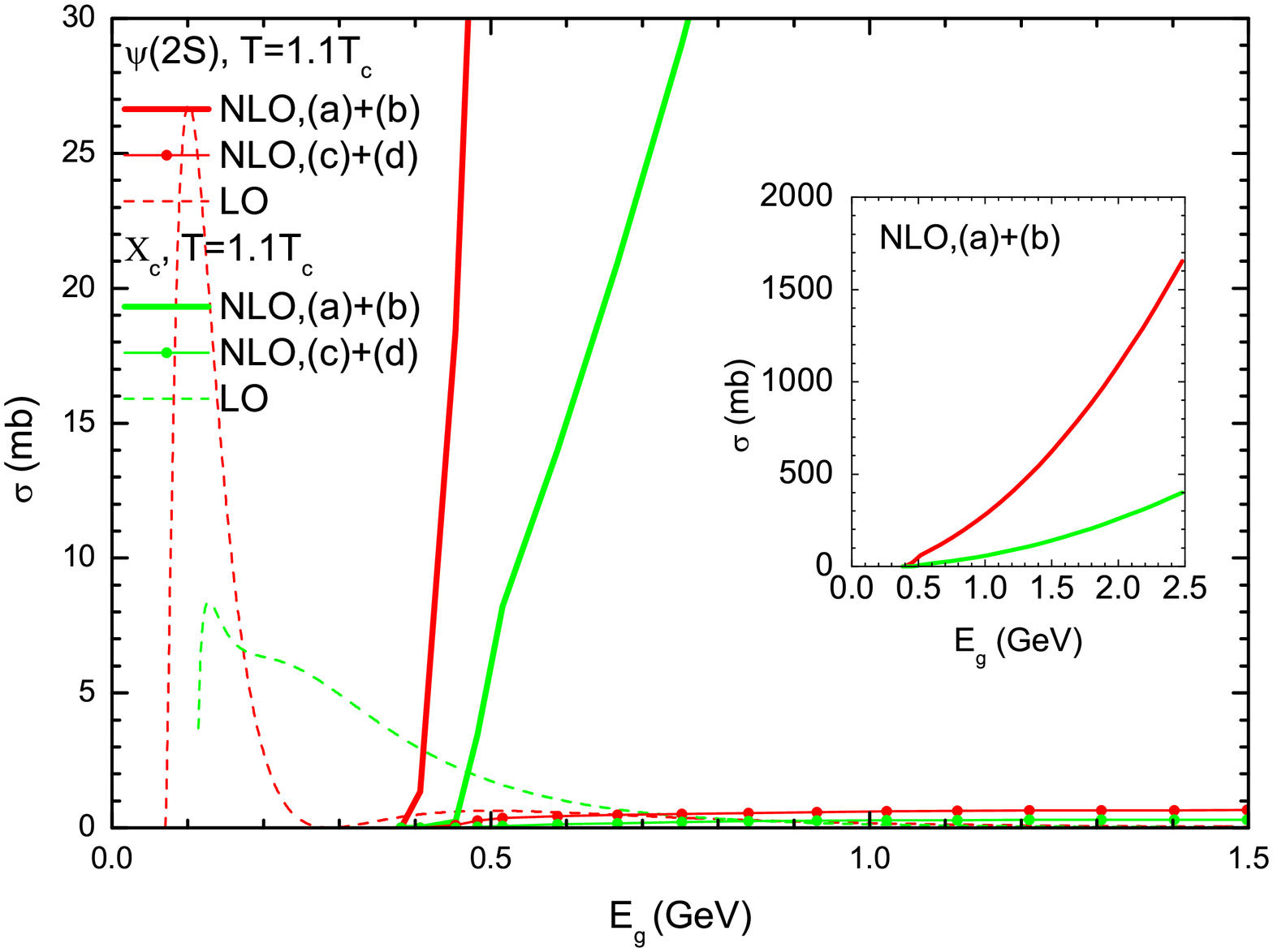}
\vspace{-0.3cm}
\caption{(Color online) NLO dissociation cross sections (from amplitudes $(a)+(b)$ and $(c)+(d)$, respectively) by thermal gluons for $J/\psi$ (upper panel),$\psi(2S)$ and $\chi_c$ (lower panel) at finite temperatures (below their respective dissociation temperature), in comparison with the corresponding LO (gluo-dissociation) cross sections. The inserts display the NLO cross sections at full scale.}
\label{fig_inmedium-charmonia-crosssections}
\end{figure}

\begin{figure} [!t]
\includegraphics[width=1.05\columnwidth]{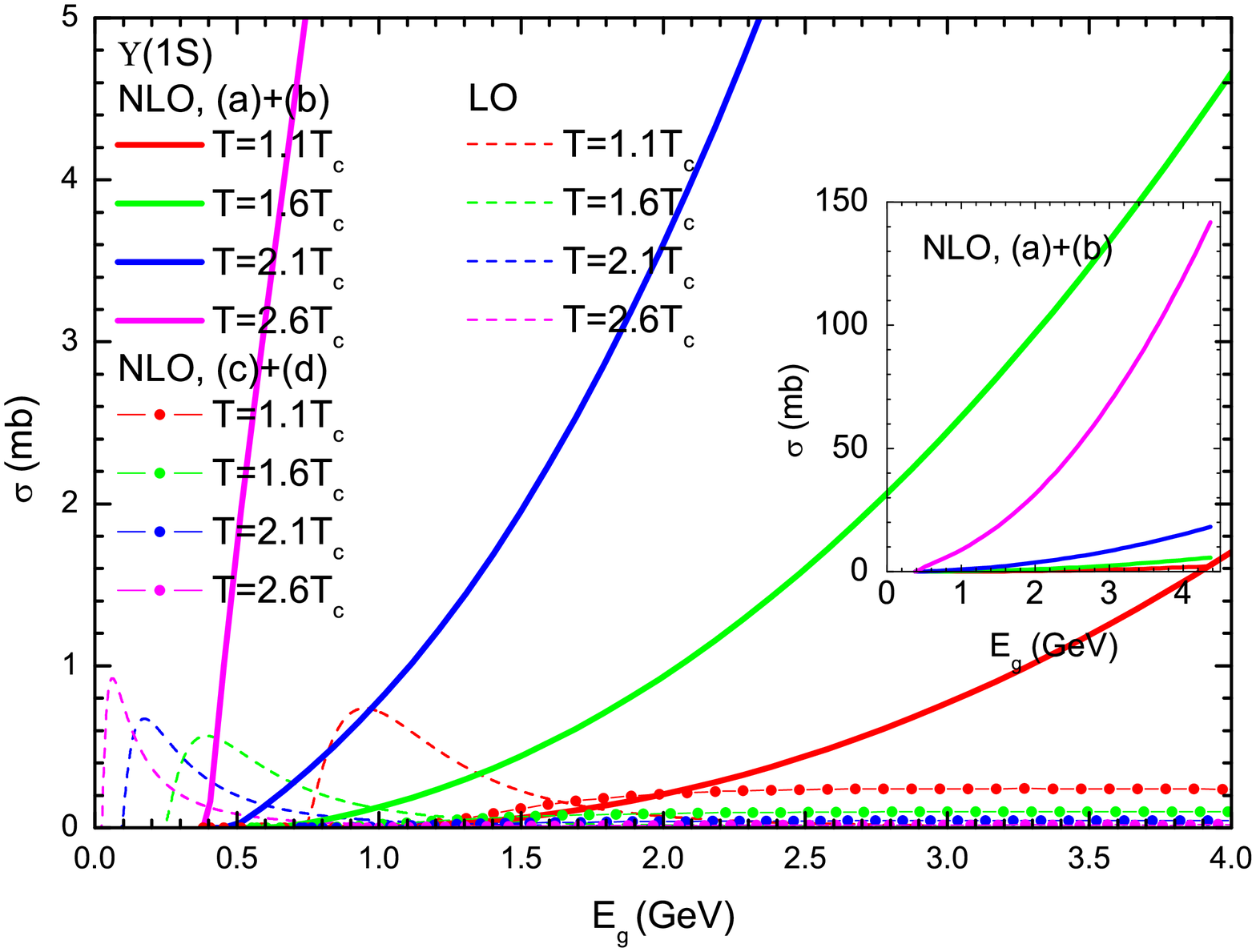}
\vspace{-0.3cm}
\includegraphics[width=1.05\columnwidth]{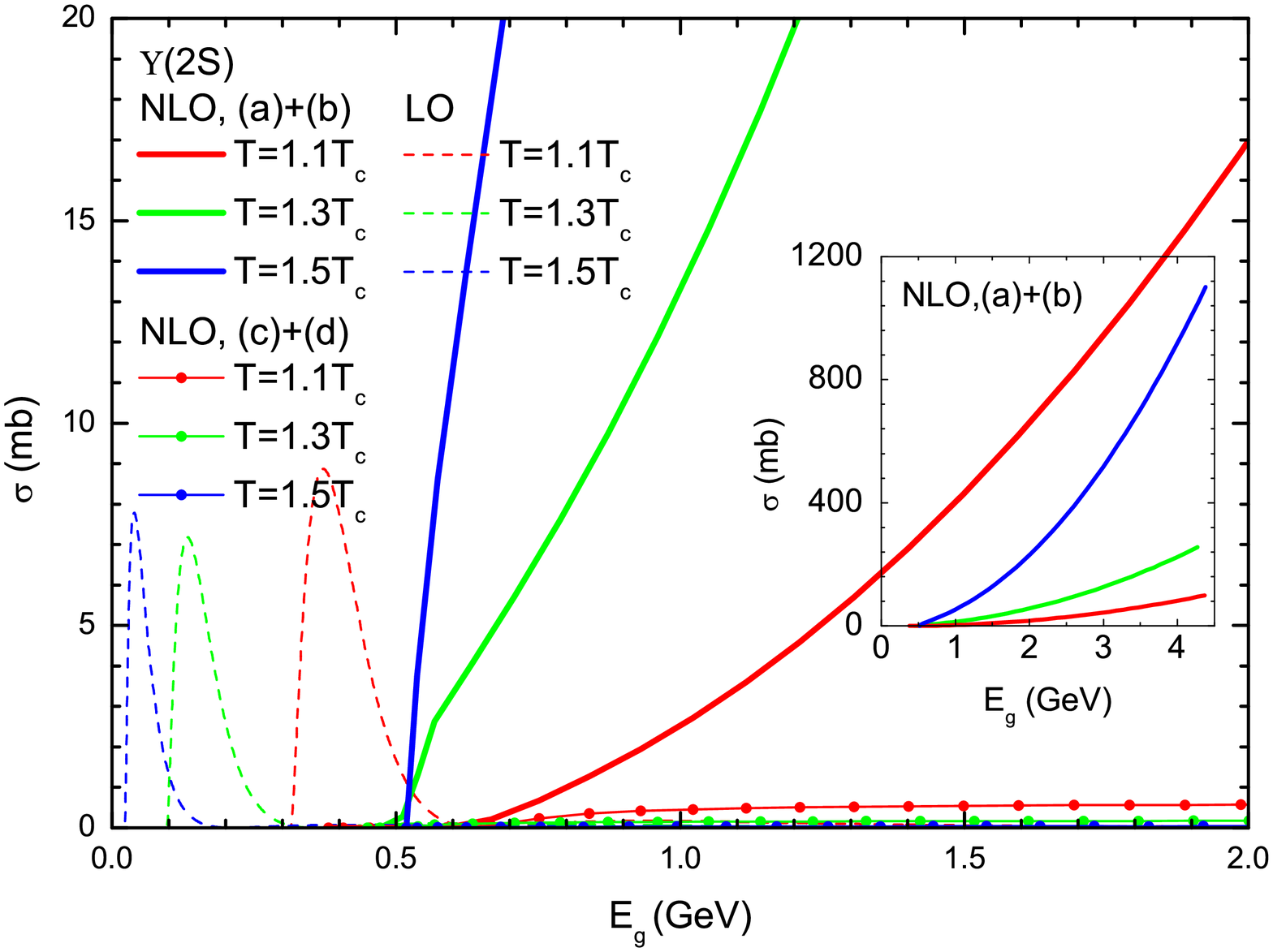}
\vspace{-0.3cm}
\includegraphics[width=1.05\columnwidth]{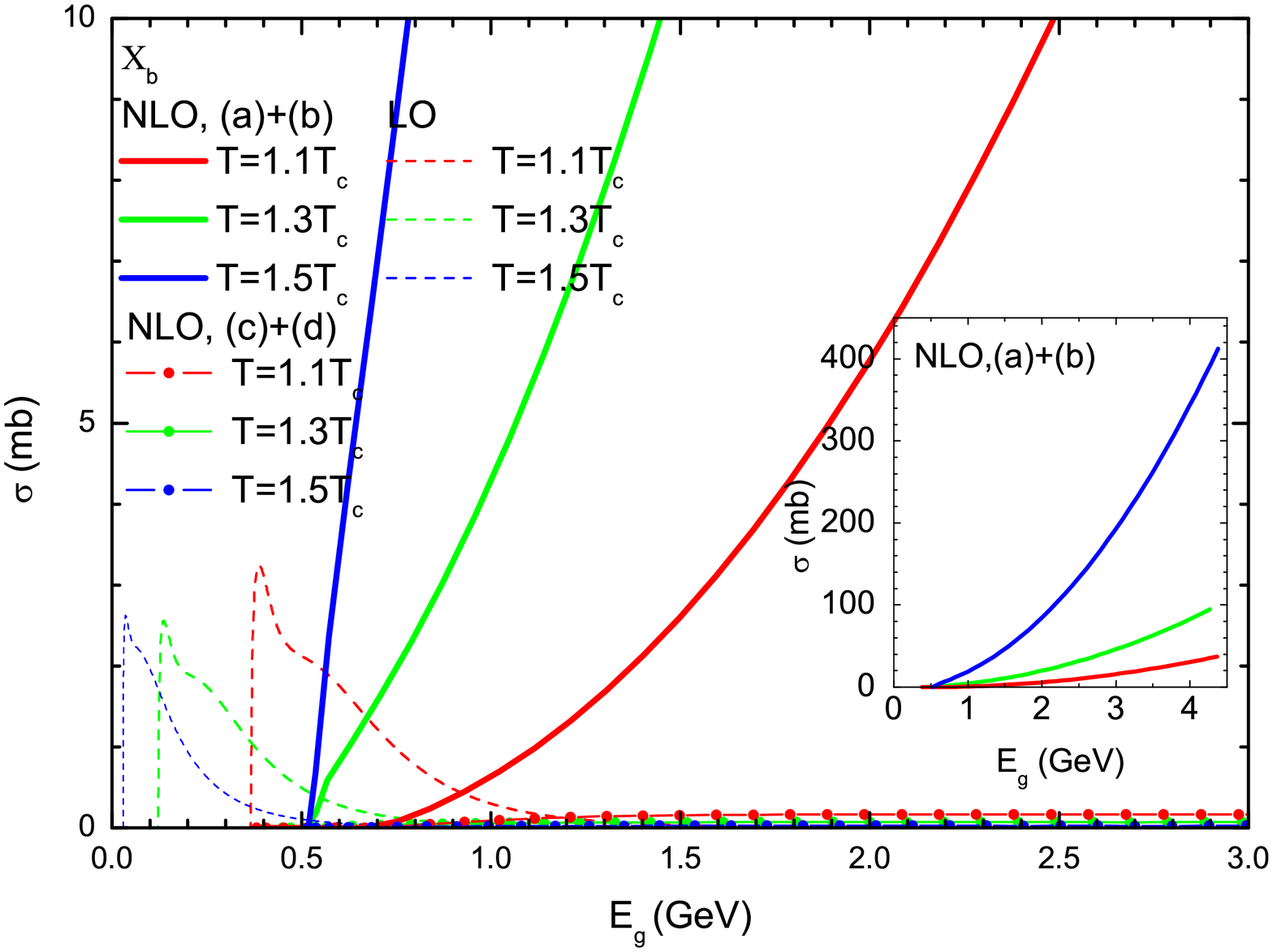}
\vspace{-0.3cm}
\caption{(Color online) Same as Fig.~\ref{fig_inmedium-charmonia-crosssections} for bottomonia: $\Upsilon(1S)$ (upper panel), $\Upsilon(2S)$ (middle panel) and $\chi_b$ (lower panel).}
\label{fig_inmedium-bottomonia-crosssections}
\end{figure}

The thus solved temperature-dependent wave functions and binding energies are then used in Eqs.~(\ref{NLOcrosssectionab}) and (\ref{NLOcrosssectioncd}) , to compute the NLO cross sections for various heavy quarkonia. The results for ground states $J/\psi$ and $\Upsilon(1S)$ in vacuum but with a ``constituent" gluon mass were already shown in Fig.~\ref{fig_vacuumcrosssections}; results of in-medium NLO cross sections for various charmonia are displayed in Fig.~\ref{fig_inmedium-charmonia-crosssections}, in comparison with the LO counterparts~\cite{Chen:2017jje}, for several temperatures below the corresponding dissociation temperature of the quarkonium under consideration. Again, unlike the LO (gluo-dissociation) cross sections dropping off toward high energies, the NLO cross sections from amplitude $(a)+(b)$ increase monotonously with the incident gluon energy, and those from amplitude $(c)+(d)$ quickly saturate at a relatively small value. The less tightly bound excited states ($\psi(2S)$ and $\chi_c$) possessing a larger radius, exhibit accordingly larger NLO as well as LO cross sections than the ground state $J/\psi$. As temperature increases, the bound state wave function expands and the binding energy decreases. While the former tends to bring up both kinds of NLO cross sections, we have checked the latter contributes oppositely (through the energy factors in Eqs.~(\ref{NLOcrosssectionab}) and (\ref{NLOcrosssectioncd})) to the temperature dependence of these two kinds of NLO cross sections. In the case of the NLO cross sections from amplitude $(c)+(d)$, the energy factor in Eq.~(\ref{NLOcrosssectioncd}) decreases with decreasing binding energy, over-counteracting the effect of expanding wave function and thus leading to an overall decrease of the cross section as temperature rises, in contrast to the fast increase of the NLO cross sections from amplitude $(a)+(b)$. When the temperature rises near to the pertinent dissociation temperature, the bound state radius grows very fast and starts to blow up~\cite{Karsch:1987pv,Chen:2017jje} and certainly overtakes the thermal gluon wave-length, thereby invalidating the dipole coupling mechanism underlying our calculations. Yet such a large NLO cross section seems to be supported by a large phenomenological dissociation rate (of the order $\sim$\,GeV) that was empirically needed in heavy quarkonium transport in QGP~\cite{Krouppa:2015yoa}. We will come back to this point later.

Similar behavior of the in-medium NLO versus LO cross sections for bottomonia has also been found, as shown in Fig.~\ref{fig_inmedium-bottomonia-crosssections}. The most tightly bound ground state $\Upsilon(1S)$ has the smallest size (and accordingly smallest cross section), guaranteeing that the dipole coupling mechanism is most applicable. Furthermore, the technical approximation made in our calculations that the rest frame of the bound state is considered also the rest frame of the final state octet $(Q\bar{Q})_8$ ({\it i.e.}, neglecting the recoil effect on the bound state by the incident gluon, see Sec.~\ref{ssec_deriveJpsiNLOcrosssecion}) should be well justified for the much more massive bottomonia. Therefore, we deem that our results of NLO cross sections are quantitatively most reliable for $\Upsilon(1S)$, especially at temperatures not too close to its dissociation temperature. The excited states $\Upsilon(2S)$ and $\chi_b$ have similar sizes and binding energies as those of $J/\psi$, and therefore, similar NLO (and also LO) cross sections, too.

%%%%%%%%%%%%%%%%%%%%%%%%%%%%%%%%%%%%%%%%%%%%%%%%%%%%%%%%%%%%%%%%
\subsection{NLO dissociation rates in QGP}
\label{ssec_dissrates}
%%%%%%%%%%%%%%%%%%%%%%%%%%%%%%%%%%%%%%%%%%%%%%%%%%%%%%%%%%%%%%%%

\begin{figure} [!t]
\includegraphics[width=1.05\columnwidth]{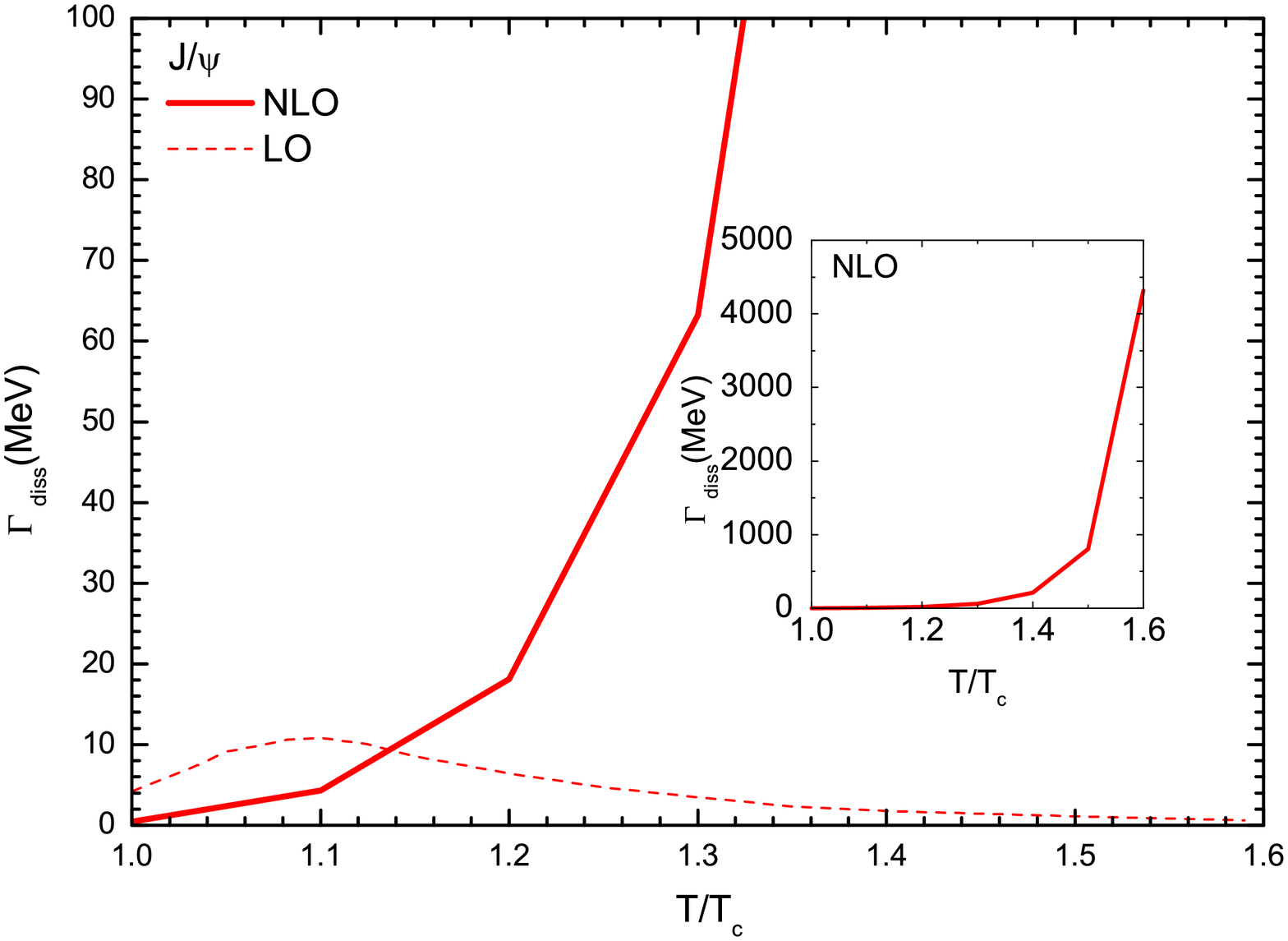}
\vspace{-0.3cm}
\includegraphics[width=1.05\columnwidth]{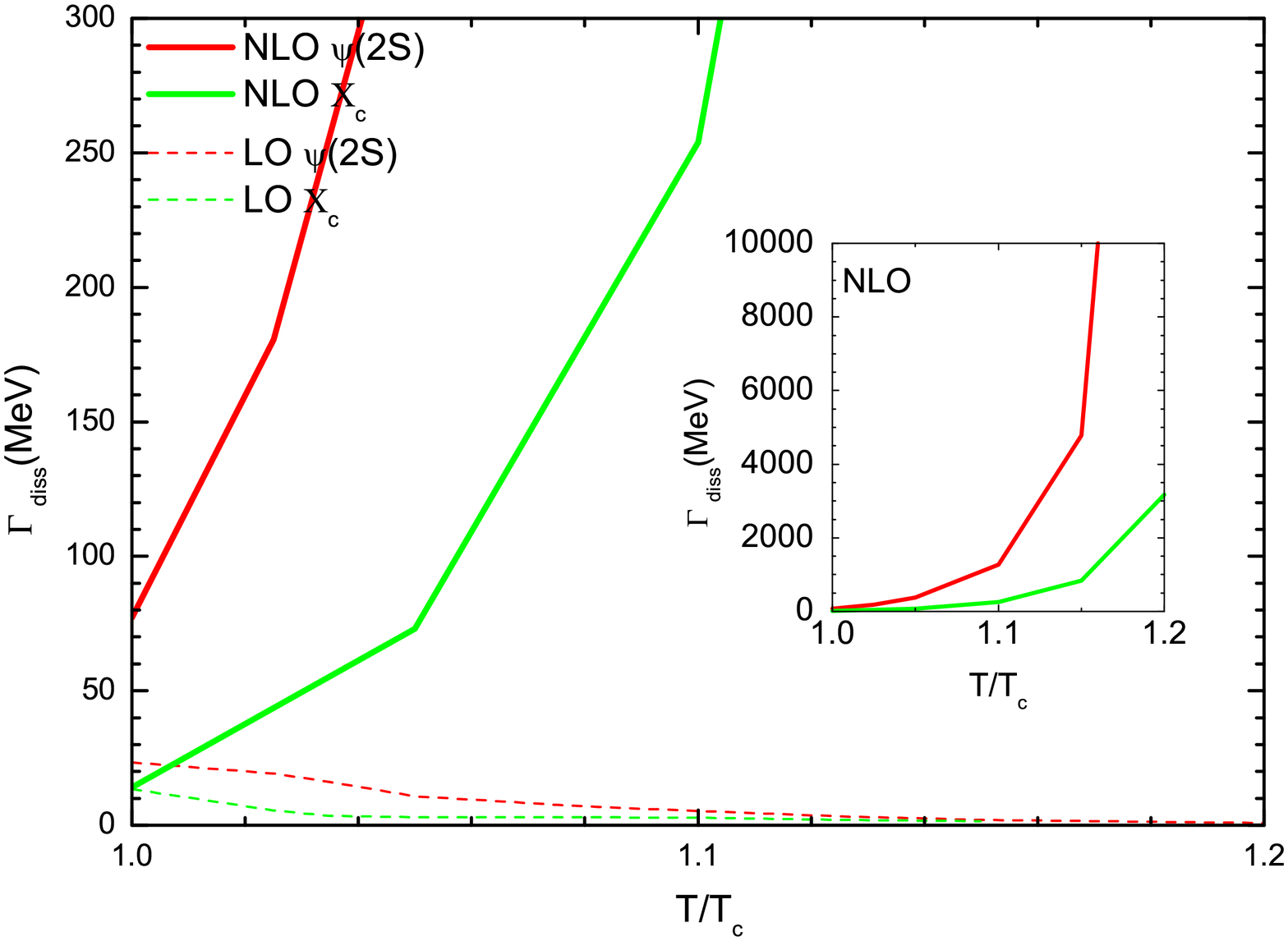}
\vspace{-0.3cm}
\caption{(Color online) Total NLO dissociation rates by thermal gluons for $J/\psi$ (upper panel),$\psi(2S)$ and $\chi_c$ (lower panel) at finite temperatures (up to their respective dissociation temperature), in comparison with the corresponding LO (gluo-dissociation) dissociation rates. The inserts display the NLO dissociation rates at full scale.}
\label{fig_charmonia_dissociation_rates}
\end{figure}

\begin{figure} [!t]
\includegraphics[width=1.05\columnwidth]{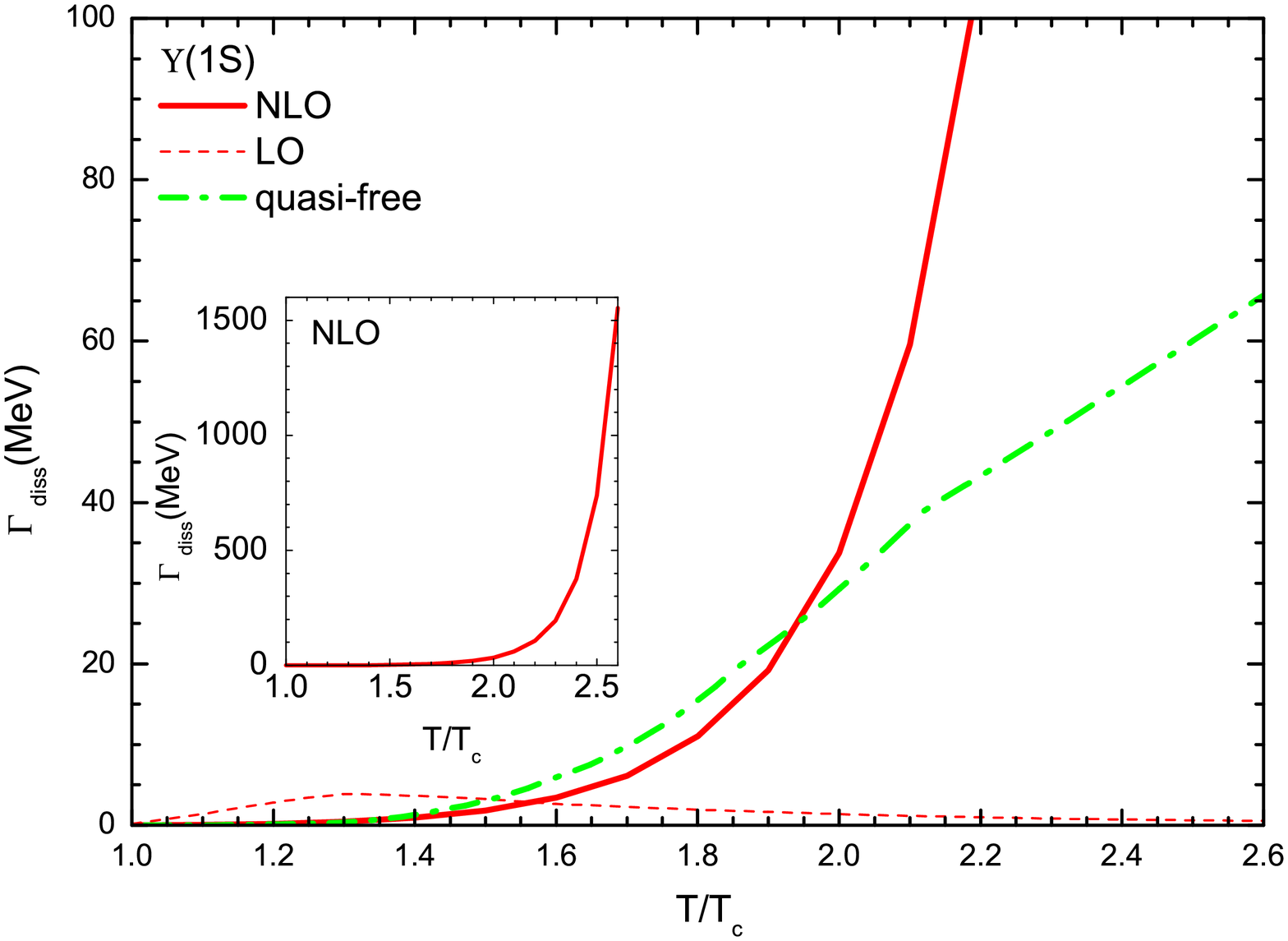}
\vspace{-0.3cm}
\includegraphics[width=1.05\columnwidth]{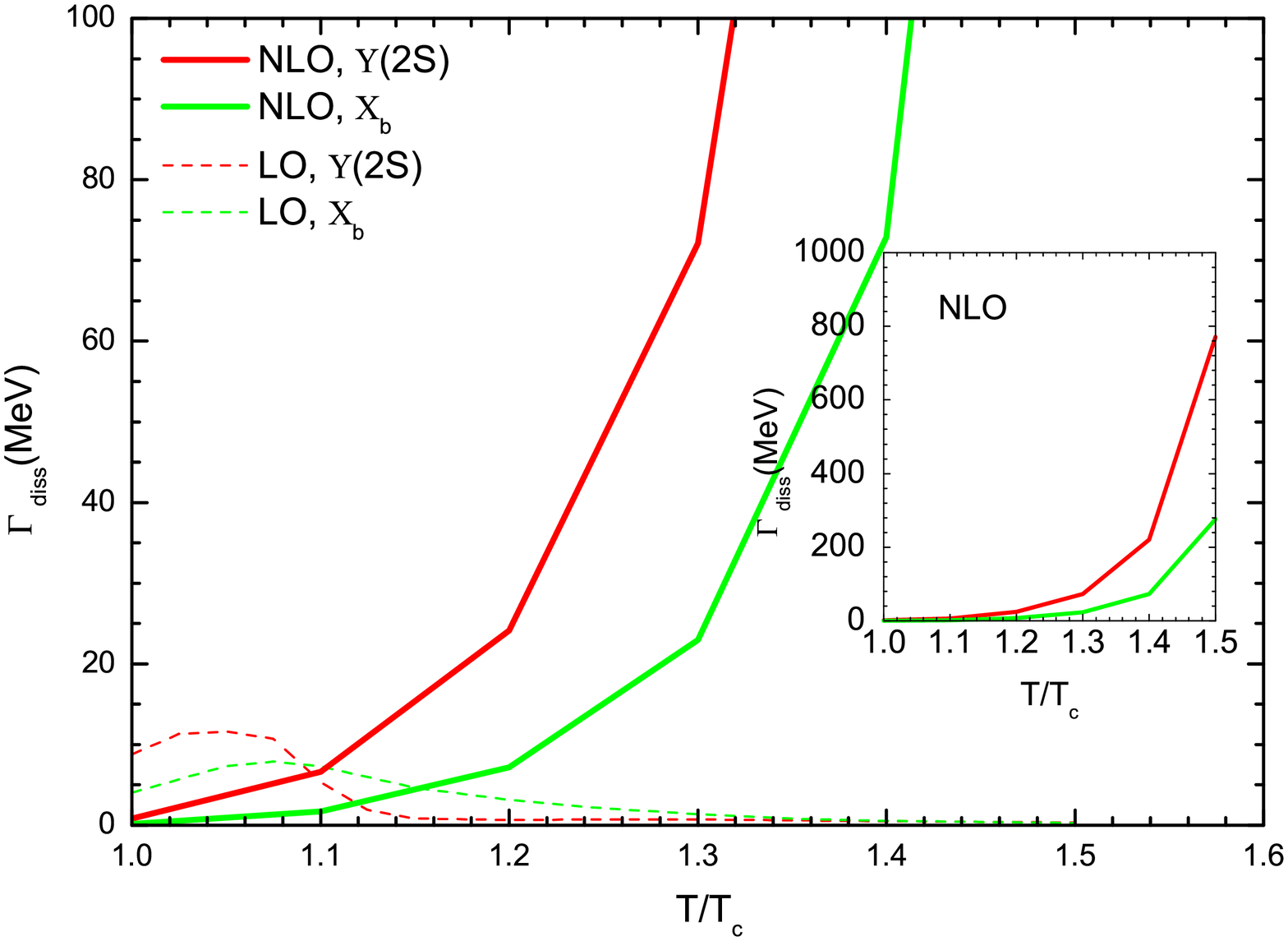}
\vspace{-0.3cm}
\caption{(Color online) Same as Fig.~\ref{fig_charmonia_dissociation_rates} for bottomonia: $\Upsilon(1S)$ (upper panel), and $\Upsilon(2S)$ and $\chi_b$ (lower panel). The quasi-free result (with gluon's contribution only) for $\Upsilon(1S)$ was taken from~\cite{Du:2017qkv}.}
\label{fig_bottomonia_dissociation_rates}
\end{figure}

The dissociation rate, an input of phenomenological studies of heavy quarkonia transport in the QGP~\cite{Rapp:2008tf,Rapp:2017chc,Zhao:2010nk,Zhou:2014kka,Strickland:2011mw,Song:2011nu}, is obtained by folding the pertinent cross section with the thermal gluon distribution. For a bound state sitting at rest in the QGP, the dissociation rate reads
\begin{equation}
\Gamma_{\rm diss}(T)=d_g\int \frac{d^3\vec{k}}{(2\pi)^3}f_g(E_g(\vec k))v_{\rm rel}\sigma(E_g(\vec k),T),
\end{equation}
where $d_g=2\cdot8=16$ is the gluon degeneracy, $v_{rel}$ the relative velocity between the incident gluon and the quarkonium at rest, and $f_g=1/(e^{E(\vec k)/T}-1)$ the Bose distribution with gluon energy $E(\vec k)=\sqrt{\vec k^2 + m_g^2(T)}$. We note that the typical thermal gluon energy is of the order of a couple of times the temperature (for massless gluons, the value is $3T$); for typical temperatures $\sim 200-400$\,MeV reached in relativistic heavy-ion collisions at RHIC and the LHC, the typical thermal gluon energy then is on the order of $\leq 1$\,GeV, which is much smaller than the mass of charmonia and bottomonia. We thus conclude that in calculating the dissociation rate, the thermal gluons do not quite probe the NLO cross section at very high energies, and therefore, the approximation we've made in calculating the NLO cross section that the recoil effect on the bound state is neglected and the rest frame of the bound state is considered also the rest frame of the final state octet $(Q\bar{Q})_8$ (see Sec.~\ref{ssec_deriveJpsiNLOcrosssecion}) is safe, especially for the more massive bottomonia. This has been verified by numerical calculations.

The calculated total NLO dissociation rates (with the total NLO cross section of the two kinds) are compiled in Fig.~\ref{fig_charmonia_dissociation_rates} and Fig.~\ref{fig_bottomonia_dissociation_rates} for charmonia and bottomonia, respectively, in comparison with the corresponding LO gluo-dissociation rates~\cite{Chen:2017jje}. At low temperatures, the bound states (especially the ground states $J/\psi$ and $\Upsilon(1S)$) are still sufficiently tightly bound, the incident gluon of long wavelength does not resolve the substructure of the bound state and therefore the LO scattering remains more effective, rendering the LO dissociation rates dominant over the NLO results. For the most tightly bound $\Upsilon(1S)$, this temperature region is relatively broad, extending up to $\sim 1.5T_c$. At higher temperatures, the NLO dissociation rates take over from the LO results that quickly drop off, and keep increasing with temperature, reaching the order of $\sim$\,GeV near the dissociation temperatures for each quarkonium. In the case of $\Upsilon(1S)$ (upper panel of Fig.~\ref{fig_bottomonia_dissociation_rates}), the calculated NLO dissociation rate was compared with the quasi-free result~\cite{Du:2017qkv}, whose underlying cross section was obtained by doubling that of the thermal gluon scattering off single bottom quark including appropriate interference effect~\cite{Du:2017qkv}. Apparently the former increases much faster than the quasi-free result toward high temperatures, which might be due to the fact that the effect of the expanding bound state wave function, while captured in the present NLO calculation, is lacking in the quasi-free scenario.

As remarked before, at temperatures very close to the heavy quarkonium dissociation temperature, the radius of bound state starts to blow up and thus the dipole coupling mechanism underlying our calculations must become invalidated. So the NLO dissociation rates near the dissociation temperatures shown here may be quantitatively questionable. Yet this large dissociation rate seems to be supported by a transport study of bottomonia phenomenology in QGP~\cite{Krouppa:2015yoa}, where, according to the authors, an empirical value of the effective ``dissociation" or ``decay"rates of order of $\geq 2$\,GeV at and above the dissociation temperature for the then unbound states was indeed needed to give a fair description of the bottomonia suppression at the LHC energy. Then if the dissociation rate were to be a continuous function across the dissociation temperature, the NLO dissociation rates of order of $\sim$\,GeV near the dissociation temperature as calculated here may not be unreasonable.

%%%%%%%%%%%%%%%%%%%%%%%%%%%%%%%%%%%%%%%%%%%%%%%%%%%%%%%%%%%%%%%%
\section{Summary}
\label{sec_sum}
%%%%%%%%%%%%%%%%%%%%%%%%%%%%%%%%%%%%%%%%%%%%%%%%%%%%%%%%%%%%%%%%

In this work, we have calculated the NLO inelastic dissociation cross sections of heavy quarkonia by thermal gluons in the QGP in the approach of second-order quantum mechanical perturbation, using a color-electric dipole coupling effective interaction Hamiltonian from the QCD multipole expansion~\cite{Yan:1980uh,Sumino:2014qpa}. Bound state effects have been systematically included and investigated, and comparisons with the LO counterparts calculated in the same framework~\cite{Chen:2017jje} have been made and examined.  The NLO cross section turns out to be divergent (infrared and soft-collinear divergence), which is regularized by thermal gluon masses in the QGP. The NLO dissociation rates obtained by folding the pertinent cross sections with the thermal gluon distribution, increases monotonously with temperature, thus removing the artifact of the LO (gluo-dissociation) approximation in which the dissociation rates decrease toward high temperatures~\cite{Chen:2017jje}. Our calculations, taking the dynamical scattering point of view, might provide another perspective of understanding the NLO break-up process of heavy quarkonium in QGP, complementing the Landau damping mechanism proposed before~\cite{Laine:2006ns,Beraudo:2007ky,Brambilla:2013dpa}. The NLO (and LO) dissociation rates evaluated in the present work, could be applied not only to the classical transport study of the Boltzmann type of the heavy quarkonium evolution in the QGP~\cite{Rapp:2008tf,Rapp:2017chc,Zhao:2010nk,Du:2017qkv,Zhou:2014kka,Strickland:2011mw,Song:2011nu}, but also might be useful in the open quantum system approach to in-medium heavy quarkonium production~\cite{Akamatsu:2011se, Brambilla:2016wgg,Blaizot:2015hya}.\\

{\bf Acknowledgments:}
We are indebted to Ralf Rapp and Xiaojian Du for useful remarks. This work was supported
by NSFC grant 11675079.


\begin{thebibliography}{99}
\bibitem{Matsui:1986dk}
  T.~Matsui and H.~Satz,
  %``$J/\psi$ Suppression by Quark-Gluon Plasma Formation,''
  Phys.\ Lett.\ B {\bf 178}, 416 (1986).
  %doi:10.1016/0370-2693(86)91404-8

\bibitem{Chatrchyan:2012lxa}
  S.~Chatrchyan {\it et al.} [CMS Collaboration],
  %``Observation of sequential Upsilon suppression in PbPb collisions,''
  Phys.\ Rev.\ Lett.\  {\bf 109}, 222301 (2012).

\bibitem{Rapp:2008tf}
  R.~Rapp, D.~Blaschke and P.~Crochet,
  %``Charmonium and bottomonium production in heavy-ion collisions,''
  Prog.\ Part.\ Nucl.\ Phys.\  {\bf 65}, 209 (2010).

\bibitem{Rapp:2017chc}
  R.~Rapp and X.~Du,
  %``Theoretical Perspective on Quarkonia from SPS via RHIC to LHC,''
  Nucl.\ Phys.\ A {\bf 967}, 216 (2017).

\bibitem{Zhao:2010nk}
  X.~Zhao and R.~Rapp,
  %``Charmonium in Medium: From Correlators to Experiment,''
  Phys.\ Rev.\ C {\bf 82}, 064905 (2010); X.~Zhao and R.~Rapp,
  %``Medium Modifications and Production of Charmonia at LHC,''
  Nucl.\ Phys.\ A {\bf 859}, 114 (2011).

\bibitem{Du:2017qkv}
  X.~Du, R.~Rapp and M.~He,
  %``Color Screening and Regeneration of Bottomonia in High-Energy Heavy-Ion Collisions,''
  Phys.\ Rev.\ C {\bf 96}, no. 5, 054901 (2017).

\bibitem{Zhou:2014kka}
  K.~Zhou, N.~Xu, Z.~Xu and P.~Zhuang,
  %``Medium effects on charmonium production at ultrarelativistic energies available at the CERN Large Hadron Collider,''
  Phys.\ Rev.\ C {\bf 89}, 054911 (2014).


\bibitem{Strickland:2011mw}
  M.~Strickland,
  %``Thermal $\upsilon_{1s}$ and chi_b1 suppression in $\sqrt{s_{NN}}=2.76$ TeV Pb-Pb collisions at the LHC,''
  Phys.\ Rev.\ Lett.\  {\bf 107}, 132301 (2011); M.~Strickland and D.~Bazow,
  %``Thermal Bottomonium Suppression at RHIC and LHC,''
  Nucl.\ Phys.\ A {\bf 879}, 25 (2012).

\bibitem{Song:2011nu}
  T.~Song, K.~C.~Han and C.~M.~Ko,
  %``Bottomonia suppression in heavy-ion collisions,''
  Phys.\ Rev.\ C {\bf 85}, 014902 (2012).

\bibitem{Peskin:1979va}
  M.~E.~Peskin,
  %``Short Distance Analysis for Heavy Quark Systems. 1. Diagrammatics,''
  Nucl.\ Phys.\ B {\bf 156}, 365 (1979); G.~Bhanot and M.~E.~Peskin,
  %``Short Distance Analysis for Heavy Quark Systems. 2. Applications,''
  Nucl.\ Phys.\ B {\bf 156}, 391 (1979).

\bibitem{Kharzeev:1994pz}
  D.~Kharzeev and H.~Satz,
  %``Quarkonium interactions in hadronic matter,''
  Phys.\ Lett.\ B {\bf 334}, 155 (1994);  D.~Kharzeev, H.~Satz, A.~Syamtomov and G.~Zinovev,
  %``On the sum rule approach to quarkonium - hadron interactions,''
  Phys.\ Lett.\ B {\bf 389}, 595 (1996).

\bibitem{Wong:2001kn}
  C.~Y.~Wong,
  %``Dissociation of heavy quarkonia in the quark gluon plasma,''
  J.\ Phys.\ G {\bf 28}, 2349 (2002); C.~Y.~Wong,
  %``Heavy quarkonia in quark-gluon plasma,''
   Phys.\ Rev.\ C {\bf 72}, 034906 (2005).


\bibitem{Arleo:2001mp}
  F.~Arleo, P.~B.~Gossiaux, T.~Gousset and J.~Aichelin,
  %``Heavy quarkonium hadron cross-section in QCD at leading twist,''
  Phys.\ Rev.\ D {\bf 65}, 014005 (2002).

\bibitem{Oh:2001rm}
  Y.~s.~Oh, S.~Kim and S.~H.~Lee,
  %``Quarkonium hadron interactions in QCD,''
  Phys.\ Rev.\ C {\bf 65}, 067901 (2002).


\bibitem{Brezinski:2011ju}
  F.~Brezinski and G.~Wolschin,
  %``Gluodissociation and Screening of $\upsilon$ States in PbPb Collisions at $\sqrt{s_{NN}}=2.76$ TeV,''
  Phys.\ Lett.\ B {\bf 707}, 534 (2012);
  %F.~Nendzig and G.~Wolschin,
  %``? suppression in PbPb collisions at energies available at the CERN Large Hadron Collider,''
  Phys.\ Rev.\ C {\bf 87}, 024911 (2013).

\bibitem{Liu:2013kkg}
  Y.~Liu, C.~M.~Ko and T.~Song,
  %``Gluon dissociation of J/¦× beyond the dipole approximation,''
  Phys.\ Rev.\ C {\bf 88}, no. 6, 064902 (2013).

\bibitem{Brambilla:2008cx}
  N.~Brambilla, J.~Ghiglieri, A.~Vairo and P.~Petreczky,
  %``Static quark-antiquark pairs at finite temperature,''
  Phys.\ Rev.\ D {\bf 78}, 014017 (2008);  N.~Brambilla, M.~A.~Escobedo, J.~Ghiglieri and A.~Vairo,
  %``Thermal width and gluo-dissociation of quarkonium in pNRQCD,''
  JHEP {\bf 1112}, 116 (2011).

\bibitem{Chen:2017jje}
  S.~Chen and M.~He,
  %``Gluo-dissociation of heavy quarkonium in the quark-gluon plasma reexamined,''
  Phys.\ Rev.\ C {\bf 96}, no. 3, 034901 (2017).

\bibitem{Grandchamp:2001pf}
  L.~Grandchamp and R.~Rapp,
  %``Thermal versus direct J / Psi production in ultrarelativistic heavy ion collisions,''
  Phys.\ Lett.\ B {\bf 523}, 60 (2001);  L.~Grandchamp, S.~Lumpkins, D.~Sun, H.~van Hees and R.~Rapp,
  %``Bottomonium production at RHIC and CERN LHC,''
  Phys.\ Rev.\ C {\bf 73}, 064906 (2006).

\bibitem{Song:2005yd}
  T.~Song and S.~H.~Lee,
  %``Quarkonium-hadron interactions in perturbative QCD,''
  Phys.\ Rev.\ D {\bf 72}, 034002 (2005); Y.~Park, K.~I.~Kim, T.~Song, S.~H.~Lee and C.~Y.~Wong,
  %``Widths of quarkonia in quark gluon plasma,''
  Phys.\ Rev.\ C {\bf 76}, 044907 (2007).

\bibitem{Laine:2006ns}
  M.~Laine, O.~Philipsen, P.~Romatschke and M.~Tassler,
  %``Real-time static potential in hot QCD,''
  JHEP {\bf 0703}, 054 (2007).

\bibitem{Burnier:2014ssa}
  Y.~Burnier, O.~Kaczmarek and A.~Rothkopf,
  %``Static quark-antiquark potential in the quark-gluon plasma from lattice QCD,''
  Phys.\ Rev.\ Lett.\  {\bf 114}, 082001 (2015).


\bibitem{Beraudo:2007ky}
  A.~Beraudo, J.-P.~Blaizot and C.~Ratti,
  %``Real and imaginary-time Q anti-Q correlators in a thermal medium,''
  Nucl.\ Phys.\ A {\bf 806}, 312 (2008).


\bibitem{Brambilla:2013dpa}
  N.~Brambilla, M.~A.~Escobedo, J.~Ghiglieri and A.~Vairo,
  %``Thermal width and quarkonium dissociation by inelastic parton scattering,''
   JHEP {\bf 1305}, 130 (2013).

\bibitem{Yan:1980uh}
  T.~M.~Yan,
  %``Hadronic Transitions Between Heavy Quark States in Quantum Chromodynamics,''
  Phys.\ Rev.\ D {\bf 22}, 1652 (1980).

\bibitem{Sumino:2014qpa}
  Y.~Sumino,
  %``Understanding Interquark Force and Quark Masses in Perturbative QCD,''
  arXiv:1411.7853 [hep-ph].


\bibitem{Brambilla:1999xf}
  N.~Brambilla, A.~Pineda, J.~Soto and A.~Vairo,
  %``Potential NRQCD: An Effective theory for heavy quarkonium,''
  Nucl.\ Phys.\ B {\bf 566}, 275 (2000);  N.~Brambilla, A.~Pineda, J.~Soto and A.~Vairo,
  %``Effective field theories for heavy quarkonium,''
  Rev.\ Mod.\ Phys.\  {\bf 77}, 1423 (2005).

\bibitem{Peskin:1995ev}
  M.~E.~Peskin and D.~V.~Schroeder,
  ``An Introduction to quantum field theory''.

\bibitem{Riek:2010fk}
  F.~Riek and R.~Rapp,
  %``Quarkonia and Heavy-Quark Relaxation Times in the Quark-Gluon Plasma,''
  Phys.\ Rev.\ C {\bf 82}, 035201 (2010);  F.~Riek and R.~Rapp,
  %``Selfconsistent Evaluation of Charm and Charmonium in the Quark-Gluon Plasma,''
  New J.\ Phys.\  {\bf 13}, 045007 (2011).

\bibitem{Karsch:1987pv}
  F.~Karsch, M.~T.~Mehr and H.~Satz,
  %``Color Screening and Deconfinement for Bound States of Heavy Quarks,''
  Z.\ Phys.\ C {\bf 37}, 617 (1988).

\bibitem{Burnier:2015tda}
  Y.~Burnier, O.~Kaczmarek and A.~Rothkopf,
  %``Quarkonium at finite temperature: Towards realistic phenomenology from first principles,''
  JHEP {\bf 1512}, 101 (2015); %``In-medium P-wave quarkonium from the complex lattice QCD potential,''
  JHEP {\bf 1610}, 032 (2016).

\bibitem{Krouppa:2015yoa}
  B.~Krouppa, R.~Ryblewski and M.~Strickland,
  %``Bottomonia suppression in 2.76 TeV Pb-Pb collisions,''
  Phys.\ Rev.\ C {\bf 92}, 061901 (2015).

\bibitem{Akamatsu:2011se}
  Y.~Akamatsu and A.~Rothkopf,
  %``Stochastic potential and quantum decoherence of heavy quarkonium in the quark-gluon plasma,''
  Phys.\ Rev.\ D {\bf 85}, 105011 (2012).


\bibitem{Brambilla:2016wgg}
  N.~Brambilla, M.~A.~Escobedo, J.~Soto and A.~Vairo,
  %``Quarkonium suppression in heavy-ion collisions: an open quantum system approach,''
  Phys.\ Rev.\ D {\bf 96}, no. 3, 034021 (2017); Phys.\ Rev.\ D {\bf 97}, no. 7, 074009 (2018).


\bibitem{Blaizot:2015hya}
  J.~P.~Blaizot, D.~De Boni, P.~Faccioli and G.~Garberoglio,
  %``Heavy quark bound states in a quark¨Cgluon plasma: Dissociation and recombination,''
  Nucl.\ Phys.\ A {\bf 946}, 49 (2016); J.~P.~Blaizot and M.~A.~Escobedo,
  %``The approach to equilibrium of a quarkonium in a quark-gluon plasma,''
  arXiv:1803.07996 [hep-ph].


\end{thebibliography}
\end{document}